\documentclass[10pt,a4paper,aps,nofootinbib,prd,twocolumn,eqsecnum,showpacs,showkeys, preprintnumbers,nofootinbib,altaffilletter]{revtex4-1}

\usepackage{hyperref}
\usepackage[english]{babel}
\usepackage{epsfig}

\usepackage[utf8]{inputenc}

\usepackage{amsmath}
\usepackage{amsfonts}
\usepackage{amssymb}

\def\be{\begin{equation}}
\def\ee{\end{equation}}

\def\bea{\begin{eqnarray}}
\def\eea{\end{eqnarray}}

\begin{document}

\title{Cosmological constrains on fast transition Unified Dark Energy/Matter models}

\author{Ruth Lazkoz$^{1}$}
\email{ruth.lazkoz@ehu.es}
\author{Iker Leanizbarrutia$^{1}$}
\email{iker.leanizbarrutia@ehu.eus}
\author{Vincenzo Salzano$^{2}$}
\email{enzo.salzano@wmf.univ.szczecin.pl}

\date{\today}

\affiliation{
${}^1$ Department of Theoretical Physics, University of the Basque Country
UPV/EHU, P.O. Box 644, 48080 Bilbao, Spain\\
${}^2$ Institute of Physics, University of Szczecin, Wielkopolska 15, 70-451 Sczcecin, Poland}

\begin{abstract}
We explore the observational adequacy of a class of  Unified Dark Energy/Matter (UDE/M) models with a fast transition. Our constraints
are set using a combination of geometric probes, some low redshift ones, and some high redshift ones (CMB related included).
The transition is phenomenologically modeled by two different transition functions corresponding to a fast and to an ultra-fast transition
respectively. We find that the key parameters governing the transition can be well constrained, and from the statistical point
of view it follows that the models cannot be discarded when compared to LCDM. We find  the intriguing result that standard/input parameters such
as $\Omega_m$ and $\Omega_b$ are far better constrained than in LCDM, and this is the case for the derived/output parameter measuring the deceleration
value at present, $q_0$.
\end{abstract}

\maketitle

\section{Introduction}

The accelerated expansion of the universe has become to be regarded as a real fact through all the high quality cosmological and astrophysical data collected
the time-period
from the pioneering
discovery announcing data release \citep{Riess:1998cb,Perlmutter:1998np} till up  to very recently \citep{Ade:2013zuv,Weinberg:2012es}. The ways  to explain
such an evolution vary from the ``simple'' addition of a $\Lambda$-term or cosmological constant \citep{Carroll:2000fy} in the general relativity (GR) framework,
to a deluge of modifications of GR itself at different levels of complexity (a very recent taxonomy can be found in \citep{Berti:2015aea}). In general, all these settings
are  usually referred to as dark  energy, see for reviews \citep{Li:2012dt,Kunz:2012aw,Copeland:2006wr,Bamba:2012cp}

On the other hand, since the analysis of the Coma galaxy cluster \citep{Zwicky:1933gu} and the discovery of the most notable properties of
rotation curves of the galaxies \citep{Rubin:1980zd}, the presence of some new kind of matter, named dark matter, who manifests itself
only through its gravitational effect, it is not only accepted, but also necessary, to explain structure formation on certain scales. Dark matter,
in the most common explanations, is supposed to be a (family of) weakly interacting particle(s), i.e. cold dark matter (CDM) under the assumption that  GR is the basic theory of gravity.
But there are also theories beyond standard GR that can mimic  the effect of dark matter without actually resorting to it, the so-called modified gravity setting.
For  dark matter reviews see \citep{Peter:2012rz,Lukovic:2014vma,Einasto:2009zd,Capozziello:2011xp}.

The effects of these two elusive components are clear and widely accepted,
 but no strong clue about their nature has been obtained so far,
and there are lots of explanations for both dark matter and dark energy.
 Normally considered as separated components of the universe,
their effects could perhaps be explained by one sole component.
In the literature we find the two most popular ways to achieve this
unification.

The first one makes the phenomenological assumption of an explicit parametrization of the equation of state in terms of the scale
factor so that the pressure/density ratio ($w=p/\rho$) has a nearly null value at the early stages of the evolution, whereas it becomes negative
enough ($w<1/3$) at late epochs.  These are usually called Unified Dark Energy/Matter (UDE/M) models. For early proposals of this sort
one can check references  \citep{Kamenshchik:2001cp,Bilic:2001cg}, whereas  \citep{Bertacca:2010ct}  reviews the topic.

The second broad idea relies on putting forward some specific coupling/interaction,
i.e. writing the respective energy conservation equations of dark matter and dark energy so that they have extra terms
which cancel each other out in the conservation equation of the sum of the two fluids.

For UDE/M models it is far from trivial how  the two components are to be separated so as to  analyze physical quantities such as the speed of sound, required
for a full fledged analysis of cosmological constraints (i.e. a study including perturbations).
In general, this is an involved problem, and, in some sense, it still remains open. Despite this difficulty, we find these models
offer an interesting arena to test whether non-conventional evolutions are admissible or even preferable.

In contrast, the coupling/interaction displays a natural separation scheme right from its inception,
and typically models in which the dark energy is a scalar field are used to model the dark energy component.
But the complexity of the equations, along with the extra number of degrees of freedom, precludes the power to reconstruct explicitly
the evolution of the background (depending on the scale factor). In fact, this is only possible if initial conditions are set, or in other words,
if a specific path is chosen in the highly multidimensional phase-space of the model. These are basically the so-called coupled quintessence scenarios,
which, despite their interest, lay beyond the main focus of this project (see \citep{Amendola:1999er} for one of the first representative contributions to the topic). But just
let us mention that recent sound observational support for some such scenarios have been presented  \citep{Salvatelli:2014zta, Valiviita:2015dfa}

UDE/M models, which are our main interest here,
 have in general been found to be somewhat inconsistent because they have to become indistinguishable from $\Lambda$CDM in order to fit the observational data \citep{Sandvik:2002jz}.
 This is the case for the generalized Chaplygin gas for example, which even in non linear evolution fails condensation to act as CDM \citep{Bilic:2003cv}. But among the different UDE/M models,
a new class has emerged that offers a way out of that difficulty: models with a fast transition between a matter-driven-like era and a dark-energy-driven-like era.
They are believed to provide an alternative and defendable explanation of the accelerated expansion of the universe
\citep{Piattella:2009kt}, as they can fit the observational data quite well while they display interesting and different new features.
Besides, fast transitions UDE/M models with scalar fields are also compatible with observational data \citep{Bertacca:2010mt}.

The theoretical reasoning on which fast transition UDE/M models rely provides hints towards well mathematically/analytically-defined expressions for their equation of state
parameter $w_{\rm UDE/M}$, as discussed in \citep{Piattella:2009kt}, but this parametrization scheme turns out to be computationally expensive when tested with likelihood techniques.
Taking into account this, and without a fundamental model, it is worth to consider simple phenomenological models for the fast transition UDE/M in order to
achieve as much theoretical progress as possible from analytical calculations.
A simple model, gentle on computations, must be one whose most important variables required for the numerical calculations can be expressed analytically.
Thus, instead of implementing the UDE/M model in the equation of state $w_{\rm UDE/M}$, as in \citep{Piattella:2009kt},
we believe a more convenient way to proceed analytically is to prescribe the evolution of the UDE/M energy density at
the level of the Hubble factor itself \citep{Bruni:2012sn}. Here we present (in section II) two parametrizations for these UDE/M set-ups
with fast transition. Then we describe  (in section III) the CMB, Galaxy Clustering and type Ia Supernovae data we use to constrain
the models. After that
we discuss (in section IV) our most relevant results, and in particular, whether the models
are statistically favoured or not as compared to the concordance scenario, $\Lambda$CDM. To close up, we present some final conclusions.

\section{UDE/M Models}

The background geometry considered for the phenomenological UDE/M models in this paper is a spatially flat Friedman-Lema\^itre-Robertson-Walker
(FLRW) metric, $ ds^2 = -dt^2 + a^2(t) \delta_{ij} dx_i dx_j $, where $a(t)$ is the scale factor as a function of the cosmic time $t$,
and $\delta_{ij}$ is the Kronecker delta. We will consider perfect fluids with densities $\rho_i$ as sources; and taking $8 \pi G=c=1 $, the Friedman equation takes the form
\be
H^2 = \left(  \dfrac{\dot{a}}{a} \right) ^2 = \sum_i \dfrac{\rho_i}{3},
\label{friedman-1}
\ee
where $H=\dot{a}/a$ is the Hubble function and the dot denotes differentiation with respect to the cosmic time. If we introduce the corresponding fractional matter-energy densities
\footnote{In the reminder, and for shortness, density will stand for fractional matter energy density.}
\be
\Omega_i(a)=\dfrac{\rho_i(a)}{3 H^2(a)} \; ,
\ee
Eq. \ref{friedman-1} becomes
\be
E^2(a)=H^2/H_0^2= \sum_i \Omega_i(a) ,
\label{friedman-2}
\ee
with $H_0$ representing the Hubble factor at present.

Being more specific about the sources, we need to clarify the role and form of our proposed UDE/M fluid: we want a UDE/M fluid which exhibits
a fast transition from the pure dark matter stage to a scenario that resembles a $\Lambda$CDM set-up. From \citep{Bruni:2012sn}, we borrow the analytical form
\be
\Omega_{UDE/M}=\Omega_t \left( \frac{a_t}{a} \right)^3 + \Omega_{\Lambda} \left[ 1 - \left( \frac{a_t}{a} \right)^3 \right] \Theta(a-a_t),
\ee
with $\Theta(a-a_t)$ playing the role of  a transition function, and $a_{t}$ the value of the scale factor at which the transition happens.
We can easily see that for $a < a_{t}$ the fluid behaves like a pure dark matter fluid, with  density $\Omega_t \left({a_t}/{a} \right)^3$. For $a > a_t$,
the fluid will rather have a density with the expression $(\Omega_t - \Omega_{\Lambda}) \left( {a_t}/{a} \right)^3 + \Omega_{\Lambda}$, thus resembling a $\Lambda$CDM scenario, as intended.

Note that, as shown in Ref \citep{Wang:2013qy}, any description for the fluid content driving the background evolution of a UDE/M setting can be mapped into that of a scenario in which dark matter and dark energy are separate components. This is so because the large-scale evolution is only sensitive to the total energy-momentum tensor, and not to the features of its separate components. But this issue should be examined under a different light if one considered perturbations, because the unified and non-unified scenarios do not have perfectly matching perturbations, as also discussed in Ref. \citep{DeSantiago:2012xh,Wang:2015wga}

Given the properties of the transition function, one can match the usual dark matter density $\Omega_{c}$\footnote{We will use the following convention:
time dependent densities will generically be expressed as $\Omega_{X}(a)$; densities evaluated now will be expressed as $\Omega_{X}$, with no additional symbols and/or suffixes.}
with the term $\Omega_t a_t^3$. Thus, the total matter component will be $\Omega_{m} = \Omega_t a_t^3 + \Omega_{b}$ when the baryonic matter term is also considered.
In principle, the Hubble factor would suffer from a degeneracy between its terms proportional to $\Omega_{c}$ and $\Omega_{b}$ if we were using only low redshift observational data.
Given that, we are going to use high redshift CMB data as well, for which this degeneracy is broken. Moreover, the use of the CMB data makes it necessary to add a radiation term,
$\Omega_{r}$ \citep{Wang2013}, which has no influence on the late-time expansion, but is fundamental in the early stages of the universe history. Thus, all in all, Eq. \ref{friedman-2} can be finally written as
\begin{eqnarray}\label{eq:friedmann-3}
E^2(a) &=& \Omega_{c} a^{-3} + \Omega_{\Lambda} \left[ 1 - \left( \frac{a_t}{a} \right)^3 \right] \Theta(a-a_t) \nonumber \\
&+& \Omega_{b} a^{-3} + \Omega_{r} a^{-4} \; .
\end{eqnarray}
 For the purpose of the statistical analysis we will perform, it is also useful to take advantage of the fact that the parameter $\Omega_{\Lambda}$
can be written as a function of the other parameters by simply evaluating  at the present time the Friedmann equation, Eq. \ref{eq:friedmann-3}, finally having
\be
\Omega_{\Lambda} = \frac{1-\Omega_{c} - \Omega_{b} - \Omega_{r}}{ ( 1 - a^{3}_t ) \Theta(a-a_t)}\, .
\label{dark-energy}
\ee

The last ingredient missing to provide a round model for this UDE/M transition is the choice of Heaviside-like functions.
We will propose two different ones; the first model for the transition will be:
\be
\Theta(a-a_t) = \frac{1}{2} +\frac{1}{\pi}\arctan \big(\beta \pi (a-a_t) \big) \;;
\label{artcan}
\ee
whereas the second transition function considered will be:
\be
\Theta(a-a_t) = \dfrac{1}{2} \left[1 + \tanh \left( 2 \beta (a-a_t) \right) \right] \; .
\label{tanh}
\ee
In both cases, the transition happens slowlier than in a  pure Heaviside function,
with the parameter $\beta$ mainly controlling the velocity of the transition;
that parameter is in fact  (and in both cases) the value of the first derivative with respect to the scale factor of the transition functions, evaluated at the transition point $a_t$.

\section{Observational data}

In this section we specify the observational data sets we have used for our analysis, and the analytical expression of the $\chi^2$ we are going to minimize in order to perform our statistical analysis.

Throughout, we will use the scale factor to redshift conversion rule
\begin{equation}
a=(1+z)^{-1}.
\end{equation}

\subsection{CMB data}

CMB data are taken from \citep{Wang2013}, where distance priors were derived from Planck first release data \citep{Ade:2013zuv}. There are two main CMB shift parameters.
The first one is
the scaled distance to photon-decoupling surface
\be
R \equiv \sqrt{\Omega_m H^2_0} \frac{r(z_*)}{c} \;.
\ee
The second one is angular scale of the sound horizon at the photon-decoupling epoch
\be
l_a \equiv \pi \frac{r(z_*)}{r_s(z_*)} \;
\ee
where $r(z_{*})$ is the comoving distance
\be
r(z_*) = \dfrac{c}{H_0} \int_0^{z_*} \frac{dz'}{E(z')}
\ee
and $r_s(z_{*})$ is the comoving sound horizon
\bea
r_s(z_*) &=&  \dfrac{1}{H_0} \int_{z_*}^\infty dz'\frac{c_s}{E(z')} \equiv \dfrac{1}{H_0} \int_0^{a_*} \frac{da'}{a'^2} \frac{c_s}{E(a')} \nonumber\\
 &=& \dfrac{c}{H_0} \int_0^{a_*} \dfrac{da'}{\sqrt{3(1+ \overline{R_b} a')a'^4 E^2(a')}} ,
\eea
where the sound speed is $c_{s} = c /\sqrt{3(1+ \overline{R_b} a)}$, with $\overline{R_b}=31500 \Omega_b h^2 (T_{CMB}/2.7K)^{-4}$, given that $T_{CMB}=2.725$ \citep{Fixsen:2009ug}, and $E(a)$ is given by Eq.~\ref{eq:friedmann-3}. Both shift parameters are evaluated at photon-decoupling epoch \citep{Hu:1995en}
\be
z_{*}= 1048 \left[ 1+0.00124(\Omega_b h^2)^{-0.738} \right] \left[ 1 + g_1 (\Omega_m h^2)^{g_2} \right] ,
\ee
where
\bea
g_1 &=& \frac{0.0783(\Omega_b h^2)^{-0.238}}{1+39.5(\Omega_b h^2)^{0.763}} \\
g_2 &=& \frac{0.560}{1+21.1(\Omega_b h^2)^{1.81}} .
\eea
The mean values and standard deviations for the triplet formed by the shift parameters and the baryon density fraction as  obtained by \citep{Wang2013} are
\bea \label{meanpluserror}
\langle l_a \rangle =  301.57 &\, ; \quad& \sigma_{l_{a}} =  0.18, \nonumber \\
\langle R \rangle = 1.7407 &\, ; \quad& \sigma_{R} =  0.0094, \\
\langle \Omega_b h^2 \rangle = 0.02228 &\ ; \quad& \sigma_{\Omega_b h^{2}} = 0.00030 \; .\nonumber
\eea
The corresponding normalized covariance matrix is:
\bea \label{Vcmb}
\bf{C_{CMB}^{norm}} = \left( \begin{array}{ccc}
1.0000 & 0.5250 & -0.4235\\
0.5250 & 1.0000 & -0.6925\\
-0.4235 & -0.6925 & 1.0000\\
\end{array} \right) \; .
\eea
In order to obtain the full covariance matrix $\bf{C_{CMB}}$, we need the transformation: $(C_{CMB})_{ij} = (C_{CMB}^{norm}) \sigma_{i} \sigma_{j} $,
where $\sigma_{i}$ are the $1\sigma$ errors of the measured best fit values given in Eq.~\ref{meanpluserror}. In order to write the CMB contribution to the $\chi^2$, we first define
\bea
\bf{X_{CMB}} = \left( \begin{array}{c}
l_a - \langle l_a \rangle \\
R - \langle R \rangle \\
\Omega_b h^2 - \langle \Omega_b h^2 \rangle \end{array} \right) \;.
\eea
and using the inverse of the covariance matrix $\bf{C_{CMB}}$, the CMB contribution to the $\chi^2$ is
\be
\chi^2_{CMB} = \bf{X_{CMB}^T C_{CMB}^{-1} X_{CMB}} \; .
\ee

\subsection{GC data}

The Galaxy Clustering (GC) data we use are the measurements of $H(z)r_s(z_d)/c$ and $D_A(z)/r_s(z_d)$ from the two dimensional two-point correlation function measured by \citep{ChuangWang2012} and \citep{Chuang:2013hya}, respectively at $z=0.35$ using the SDSS DR7 Luminous Red Galaxies sample \citep{Kazin:2009cj}, and at $z=0.57$ using the CMASS galaxy sample from BOSS \citep{Eisenstein:2011sa}. Here, $H(z)$ is the Hubble function defined in Eq.~\ref{friedman-2}~-~\ref{eq:friedmann-3}; and $D_A(z)$ is the angular diameter distance

The Galaxy Clustering (GC) data we use are the measurements of $H(z)r_s(z_d)/c$ and $D_A(z)/r_s(z_d)$ obtained by \citep{Wang2013}
from the two dimensional two-point correlation function measured at  $z=0.35$ using the SDSS DR7 Luminous Red Galaxies sample \citep{ChuangWang2012},
and at $z=0.57$ using the CMASS galaxy sample from BOSS  \citep{Chuang:2013hya}. Here, $H(z)$ is the Hubble function defined in
Eqs.~\ref{friedman-2}~-~\ref{eq:friedmann-3}; and $D_A(z)$ is the angular diameter distance:
\be
D_A(z) = \frac{c}{1+z} \int_0^z \frac{dz'}{H(z')} \; .
\ee
The comoving sound horizon $r_s(z_d)$ is evaluated at the dragging epoch \citep{Eisenstein:1997ik}
\be
z_d = \dfrac{1291(\Omega_m h^2)^{0.251}}{1+0.659(\Omega_m h^2)^{0.828}} \left[ 1+ b_1 (\Omega_b h^2)^{b_2} \right] \; ,
\ee
with
\bea
b_1 &=& 0.313(\Omega_m h^2)^{-0.419} \left[ 1+0.607(\Omega_m h^2)^{0.674} \right] \; \; \\
b_2 &=& 0.238(\Omega_m h^2)^{0.223} \; .
\eea
The mean values and the $1 \sigma $ errors for these quantities are
\bea\label{Vgc1}
\left\langle \frac{H(0.35)r_s(z_d)}{c} \right\rangle &=& 0.0434 \pm 0.0018 \nonumber \\
\left\langle \frac{D_A(0.35)}{r_s(z_d)} \right\rangle &=& 6.60 \pm 0.26  \; ,
\eea
with normalized correlation coefficient $r_{0.35} = 0.0604$; and
\bea\label{Vgc2}
\left\langle \frac{H(0.57)r_s(z_d)}{c} \right\rangle &=& 0.0454 \pm 0.0031 \nonumber \\
\left\langle \frac{D_A(0.57)}{r_s(z_d)} \right\rangle &=& 8.95 \pm 0.27
\eea
with normalized correlation coefficient $r_{0.57}=0.4874$. The GC contribution is calculated independently for each redshift, $\chi^2_{GC}= \chi^2_{GC1} + \chi^2_{GC2}$, where each term is
\be
\chi^2_{GCi} = \frac{1}{1-r_i^2} \left( \frac{X1^2_{GCi}}{\sigma_{1i}^2} + \frac{X2^2_{GCi}}{\sigma_{2i}^2} - 2 r_i \frac{X1_{GCi}}{\sigma_{1i}} \frac{X2_{GCi}}{\sigma_{2i}} \right) \, , \\
\ee
where $r_i$ is the correlation between the two functions at each redshift, and
\bea
  X1_{GCi} &=& \frac{H(z_i)r_s(z_d)}{c}  - \left\langle \frac{H(z_i)r_s(z_d)}{c} \right\rangle
  , \; \; \\
  X2_{GCi} &=& \frac{D_A(z_i)}{r_s(z_d)} - \left\langle \frac{D_A(z_i)}{r_s(z_d)} \right\rangle \, .
\eea

\subsection{SNe Ia data}

The SNe Ia dataset we have used is the Union2.1 compilation \citep{Suzuki:2011hu}, made of $580$ Type Ia Supernovae distributed in the redshift interval $0.015 < z < 1.414$. The dataset provides the distance modulus $\mu(z_i)$ for each SN and the full statistical plus systematics covariance matrix. The distance modulus is defined as
\be
\mu(z) = 5 \log_{10} d_L(z) + \mu_{0} \;,
\ee
where $d_{L}$ is the dimensionless luminosity distance given by
\be
d_L(z) = (1+z) \int_0^z \frac{dz'}{E(z')} ,
\ee
and $\mu_{0}$ is a nuisance parameter involving the value of the Hubble constant $H_{0}$ and the SNeIa absolute magnitude. Defining the difference vector between the model and the observed magnitudes
\bea
\bf{X_{SN}} = \left( \begin{array}{c}
\mu(z_{1}) - \mu_{obs}(z_{1}) \\
\ldots \\
\mu(z_{\mathcal{N}}) - \mu_{obs}(z_{\mathcal{N}})
\end{array} \right) \; ,
\eea
and using the covariance matrix $\bf{C}$ given by \citep{Suzuki:2011hu}, we could build
\be
\chi^2_{SN}=\bf{X_{SN}}^T \cdot C^{-1} \cdot \bf{X_{SN}} .
\ee
However, this $\chi^2_{SN}$ expression would contain the nuisance parameter $\mu_{0}$. In order to get rid of the parameters degeneracy intrinsic to its definition, we marginalize over $\mu_{0}$. In such case it is quite easy to perform the marginalization analytically, because the parameter $\mu_{0}$ enters the definition of the observational quantities as an additive constant. Moreover, we are assuming, as usual, a flat non-informative prior on it. For the sake of clarity, we do not report all the algebraic steps which are required to obtain the final expression of the marginalized $\chi^2$; for the interested readers, all details are in Appendix C of \citep{Conley:2011ku}. The SNeIa $\chi^2$ contribution, after marginalizing over $\mu_{0}$, is:
\be
\chi^2_{SN} = a + \log \frac{d}{2 \pi } - \frac{b^2}{d} \, ,
\ee
where: $a \equiv  {\bf X_{SN}}^{T} \cdot {\bf C}^{-1} \cdot  \bf{X_{SN}}$, $b \equiv  {\bf X_{SN}}^{T} \cdot {\bf C} ^{-1} \cdot \bf{1}$, and $d \equiv {\bf 1}^{T} \cdot {\bf C}^{-1} \cdot \bf{1}$, with $\bf{1}$ standing for the identity matrix.

A Gaussian prior for the Hubble constant is also assumed: $H_0 = 100 \: h$ km s$^{-1}$ Mpc$^{-1}$ $= 69.6 \pm 0.7$ km s$^{-1}$ Mpc$^{-1}$ \citep{Bennett:2014tka};
thus, its contribution to the total $\chi^2$ is the following:
\be
\chi^{2}_{H_{0}} = (100 \, h - H_0)^2/\sigma_{H_0}^2 .
\ee
The total $\chi^2$ will be, of course, the sum of all the terms previously described:
\be
\chi^2 = \chi^{2}_{CMB} + \chi^{2}_{GC} + \chi^{2}_{SN} + \chi^{2}_{H_{0}}
\ee

\section{Results and Discussion}

{\renewcommand{\tabcolsep}{1.mm}
{\renewcommand{\arraystretch}{1.75}
\begin{table*}[htbp]  
\begin{minipage}{\textwidth}
\caption{Summary of constraints. Median values for the free parameters and the corresponding value for the parameters $\Omega_{\Lambda}$ and $q_0$, using CMB, GC and SN data. The value for the minimum $\chi^2_{red}$ and the Bayesian ratios with respect to $\Lambda$CDM are also shown.}\label{tablaSN}
\centering
\resizebox*{\textwidth}{!}{
\begin{tabular}{ccccccccccc}
\hline
\bf{Model} & & $h$ & $\bf{\Omega_c}$ & $\bf{\Omega_b}$ & \bf{parameter $\sharp 4$} & $\beta$ & $\bf{\Omega_{\Lambda}}$ & $\bf{q_0}$ & $\bf{\chi^2_{red}}$ & $\bf{ln B_{i\Lambda}}$ \\
\hline \hline
arctan & & $0.69508^{+0.00068}_{-0.00064}$ & $0.2445^{+0.0011}_{-0.0012}$ & $0.04626^{+0.00024}_{-0.00026}$ & $a_t = 0.170^{+0.010}_{-0.011}$ & $552^{+75}_{-69}$ & $0.7091^{+0.0015}_{-0.0013}$ & $-0.56907^{+0.00099}_{-0.00120}$ & $0.9501$ & $+0.791$\\
tanh & & $0.69587^{+0.00065}_{-0.00061}$ & $0.2436^{+0.0011}_{-0.0011}$ & $0.04597^{+0.00023}_{-0.00024}$ & $a_t = 0.1809^{+0.0099}_{-0.0105}$ & $771^{+71}_{-67}$ & $0.7103^{+0.0014}_{-0.0013}$ & $-0.57181^{+0.00085}_{-0.00103}$ & $0.9498$ & $+0.902$\\
$\Lambda$CDM & & $0.6906_{-0.0093}^{+0.0090}$ & $0.245^{+0.011}_{-0.010}$ & $0.04692^{+0.00097}_{-0.00096}$ & $-$ & $-$ & $0.708_{-0.012}^{+0.011}$ & $-0.562_{-0.017}^{+0.017}$ & $0.9488$ & $0$\\
quiessence & & $0.696_{-0.017}^{+0.016}$ & $0.243^{+0.012}_{-0.012}$ & $0.0461^{+0.0024}_{-0.0022}$ & $w=-1.029^{+0.062}_{-0.067}$ & $-$ &  $0.712_{-0.015}^{+0.014}$ & $-0.597_{-0.087}^{+0.085}$ & $0.9503$ & $-0.242$\\
\hline \hline
\end{tabular}}
\end{minipage}
\end{table*}}}

The statistical analysis will be performed by minimizing the $\chi^2$ function using the Markov Chain Monte Carlo (MCMC) Method \cite{Christensen:2001gj,Lewis:2002ah,Trotta:2004qj}. The statistical convergence of each MCMC round has been tested using the method
described in \cite{Dunkley2005}. In order to state the effective statistical weight and validity of our UDE/M models, we have also analysed the $\Lambda$CDM model: \citep{Carroll:1991mt,Sahni:1999gb}
\be
E_{\Lambda CDM}^2(z) = (\Omega_c + \Omega_b)a^{-3}+\Omega_{r}a^{-4 }+ \Omega_{\Lambda} \, ,
\ee

and the quiessence model \citep{Knop:2003iy,Riess:2004nr}
\be
E_{Q}^2(z) = (\Omega_c + \Omega_b)a^{-3}+\Omega_{r}a^{-4 } + \Omega_{\Lambda}a^{-3(1+w)}.
\ee
In both cases, the usual matter density $\Omega_m$ has been split into the dark matter $\Omega_c$ and the  baryonic $\Omega_b$  densities terms, in order to have the same parameters as the UDE/M models.

The priors on the parameters that we have chosen are as general as possible: a positive dark matter density between $0<\Omega_{c}<1$; a positive baryonic matter density smaller than the dark matter density $0 < \Omega_b < \Omega_{c}$; a positive Hubble function $ E(a) > 0 $ for all $a$ values; and $0 < a_t < 1 $ because we want
the transition to actually have happened.

\begin{figure}[b]
\begin{minipage}{0.45\textwidth}
 \centering
  \includegraphics[width=0.48\textwidth]{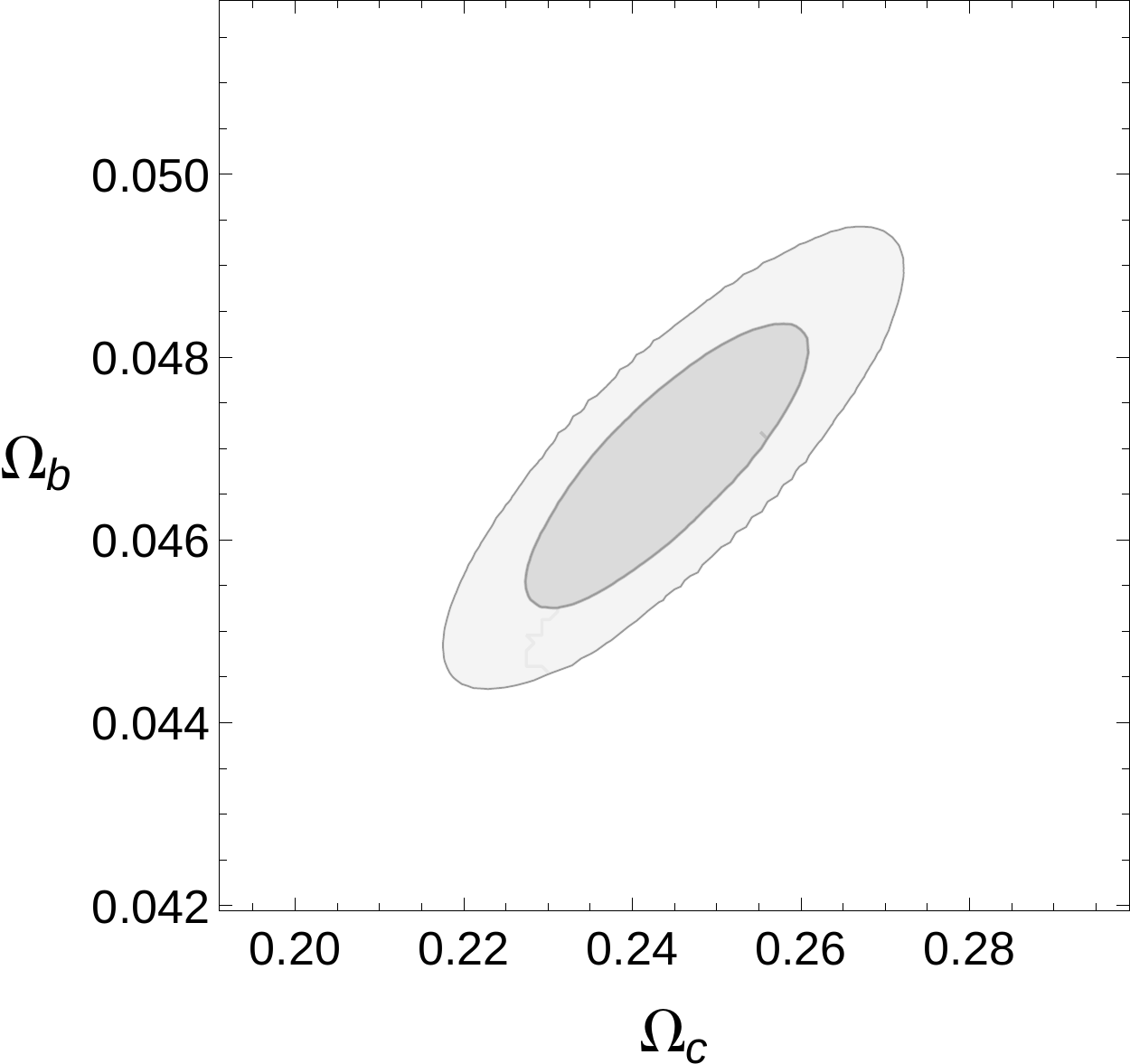}~~
  \includegraphics[width=0.45\textwidth]{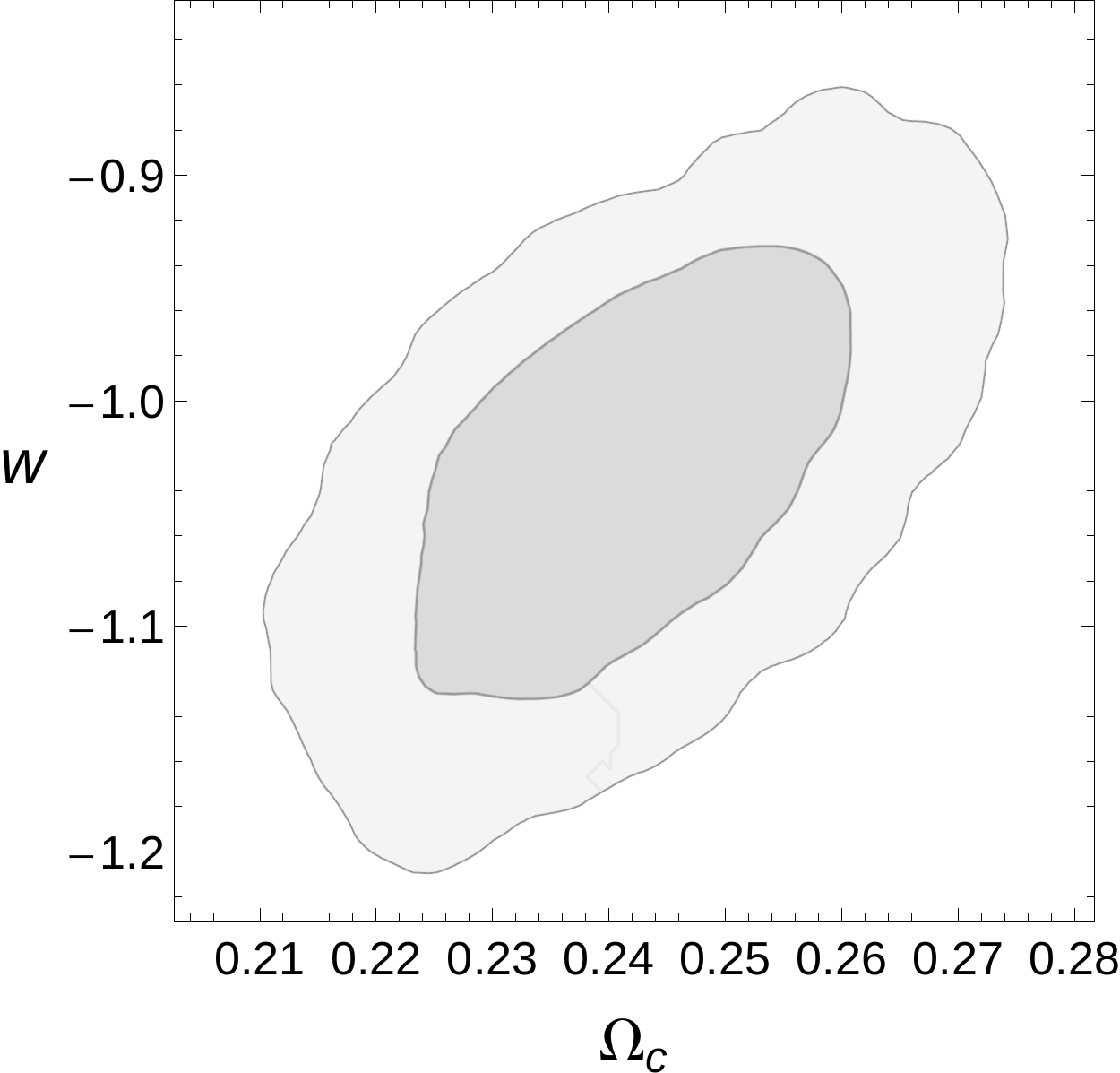}
\caption{Contour plot for the and $\Lambda$CDM (left) and quiessence (right) models, dark grey areas are $1\sigma$ region and light grey areas are $2\sigma$ region.} \label{fig:LCDM}
\end{minipage}
\end{figure}

Fig.~\ref{fig:LCDM} shows confidence regions for the $\Lambda$CDM and quiessence models, while a summary of the results of our statistical analysis
can be found in Table~\ref{tablaSN} where the reduced best-fit $\chi^2$ is also shown. However, more robust conclusions for the model selection can be drawn only
from the Bayes factors after having computed the statistical evidence. For the computation of the evidence we have used the nested sampling algorithm \citep{Mukherjee:2005wg}.
The Bayes factor \citep{Trotta:2005ar} is obtained comparing the UDE/M and quiessence models with the $\Lambda$CDM,
assumed as reference model, and the results are shown in the last column of Table~\ref{tablaSN}.

\begin{figure*}[htbp]
\begin{minipage}{1.0\textwidth}
 \centering
  \includegraphics[width=0.3\textwidth]{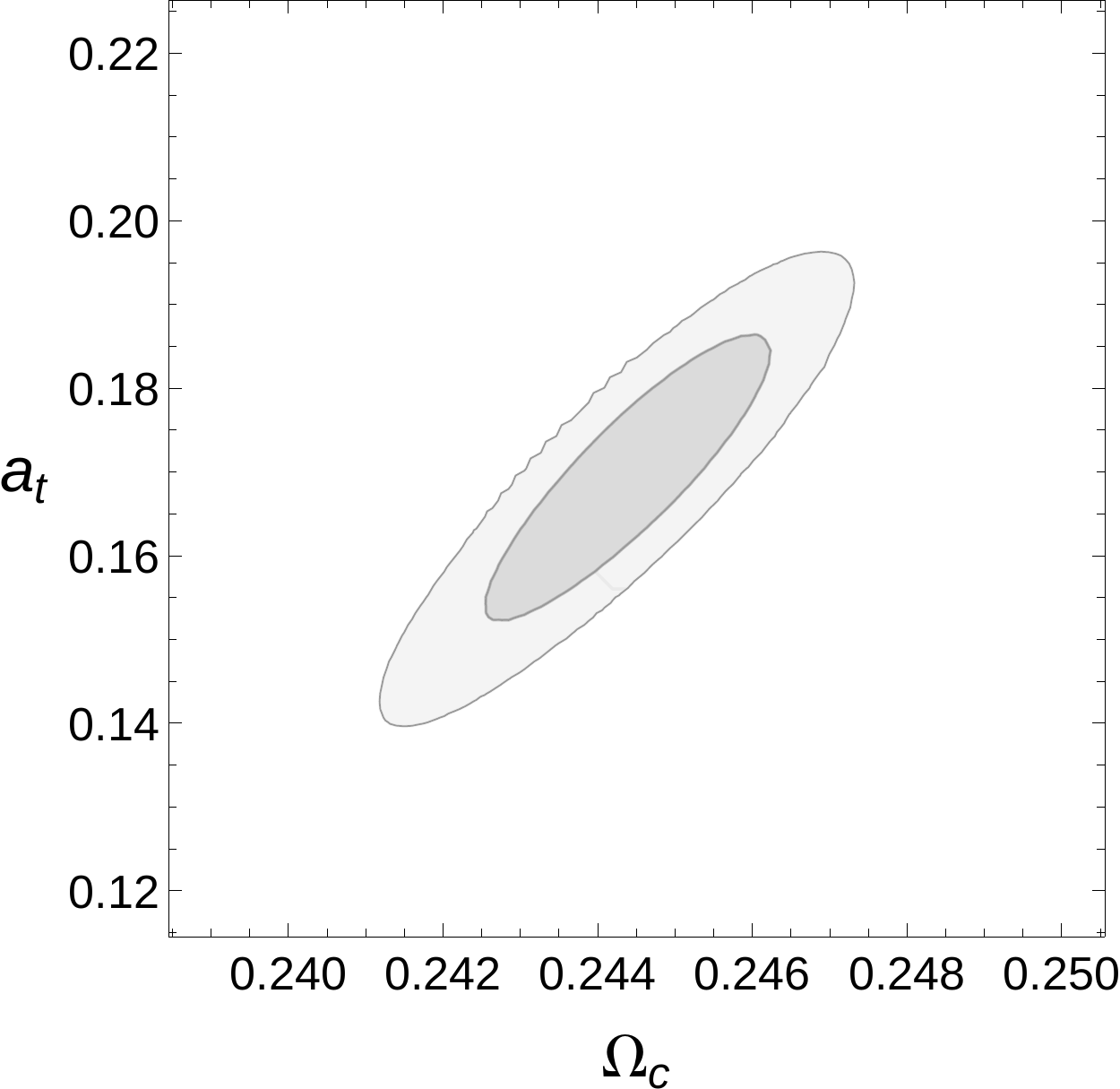}~~~~~
  \includegraphics[width=0.3\textwidth]{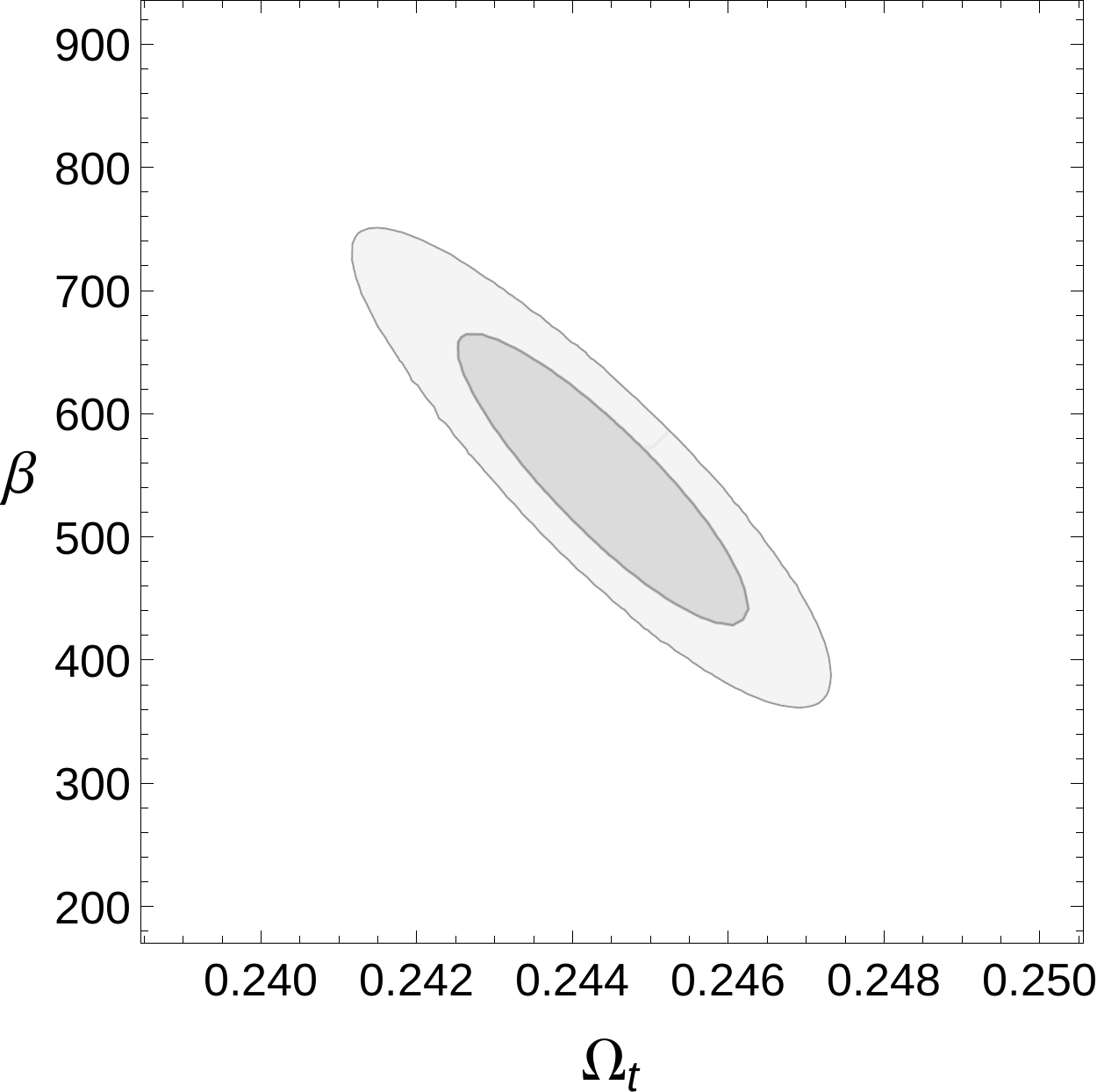}~~~~~
  \includegraphics[width=0.3\textwidth]{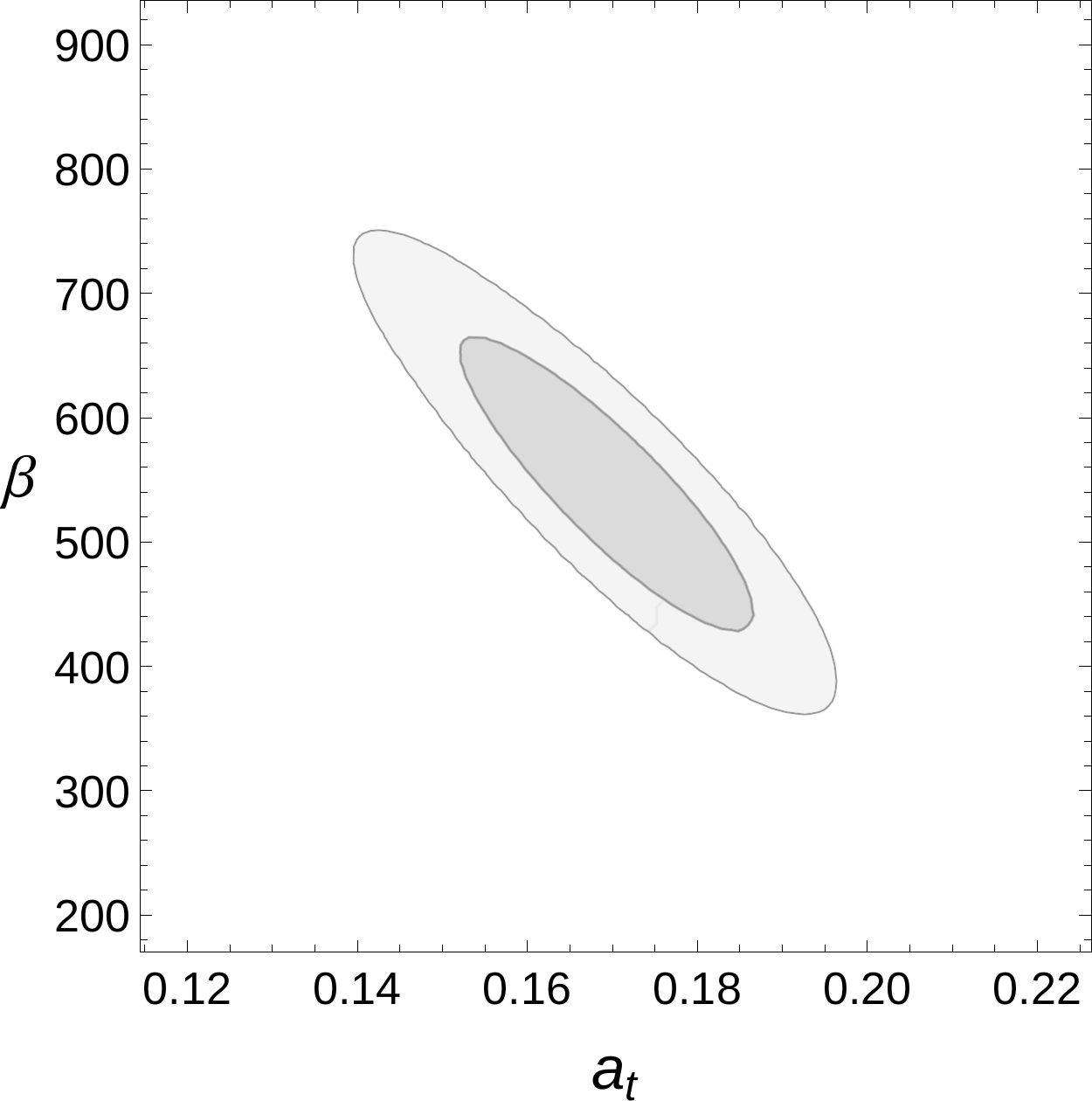}\\
  ~~~\\
  \includegraphics[width=0.3\textwidth]{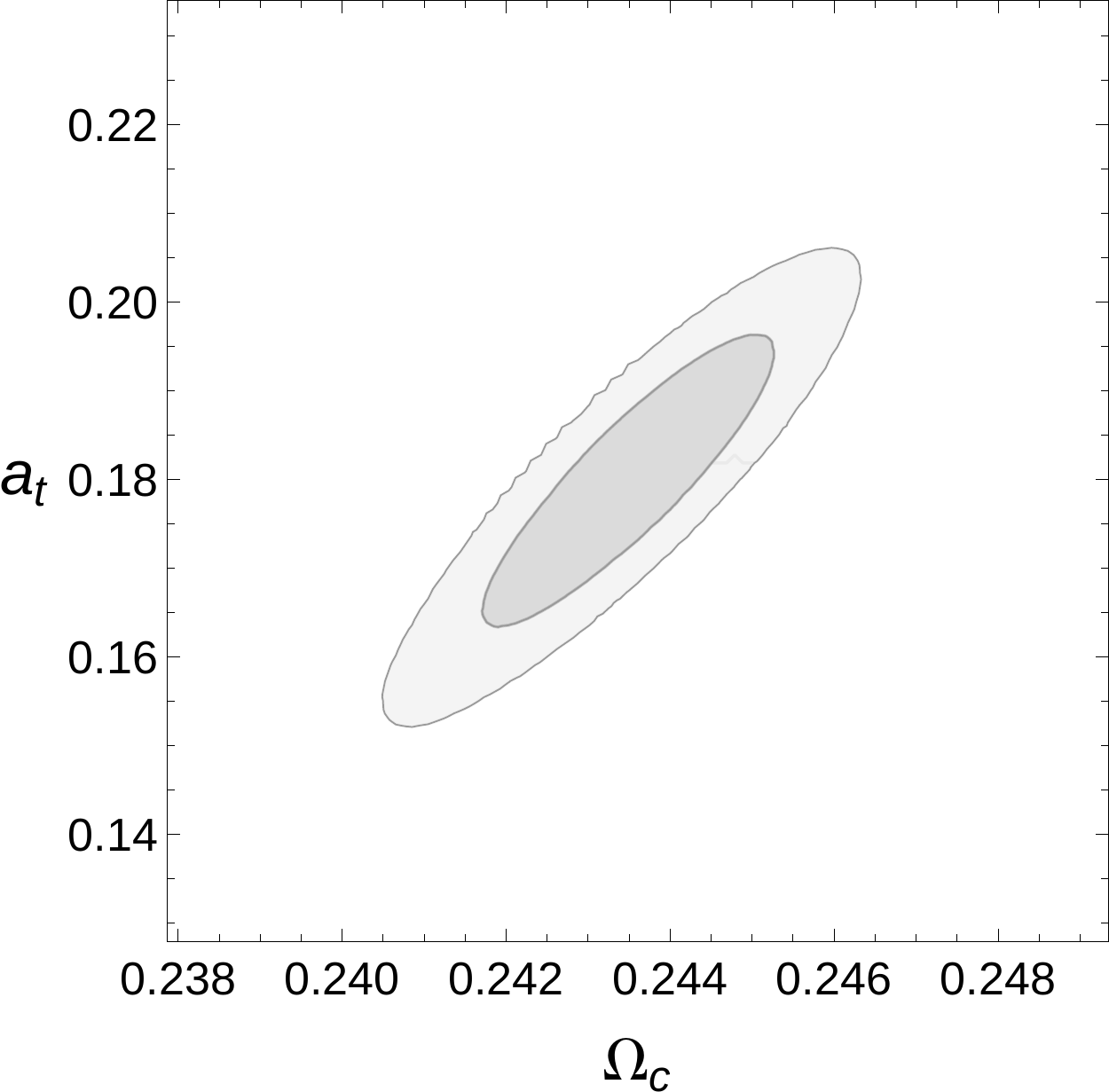}~~~~~
  \includegraphics[width=0.3\textwidth]{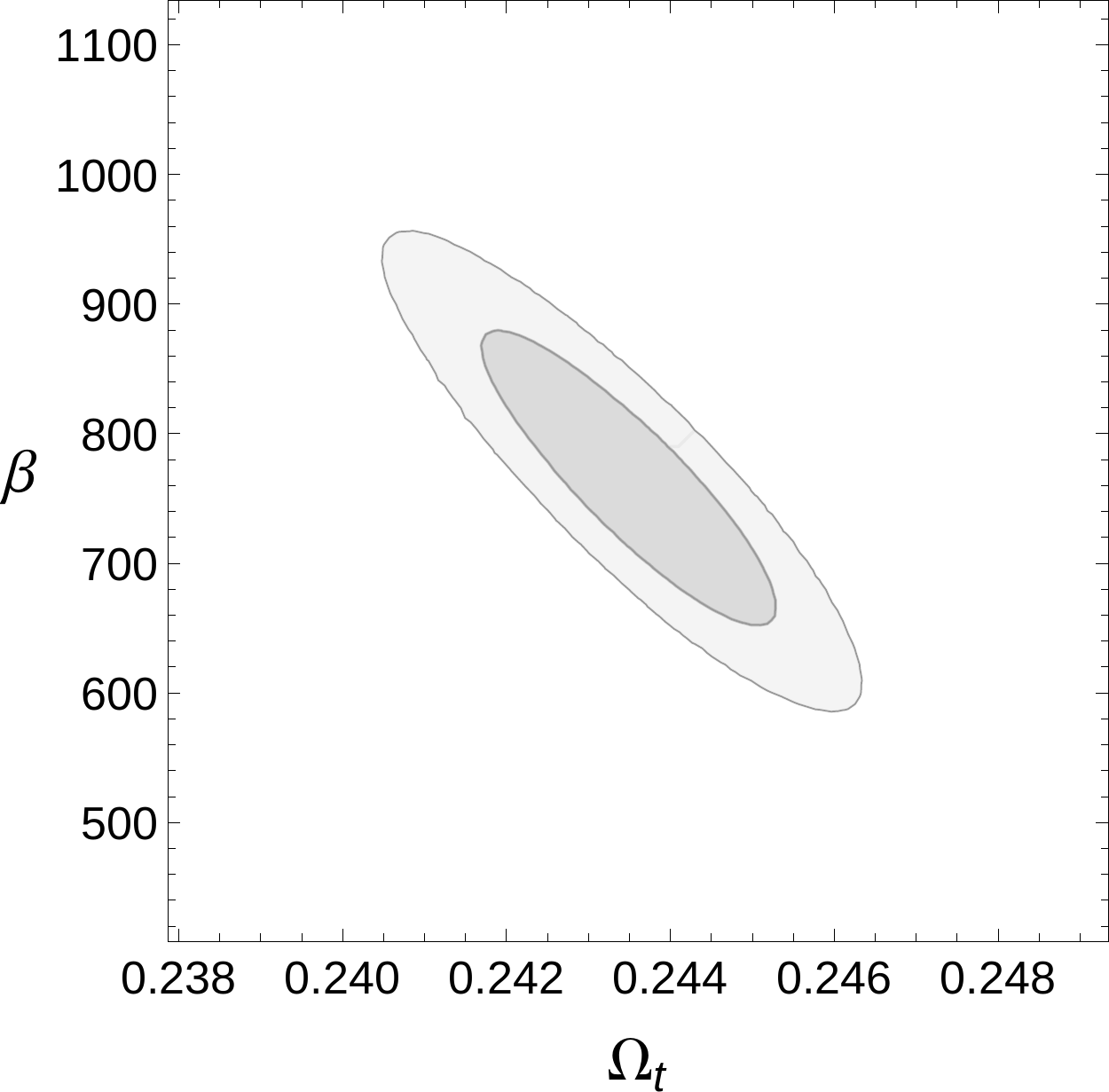}~~~~~
  \includegraphics[width=0.3\textwidth]{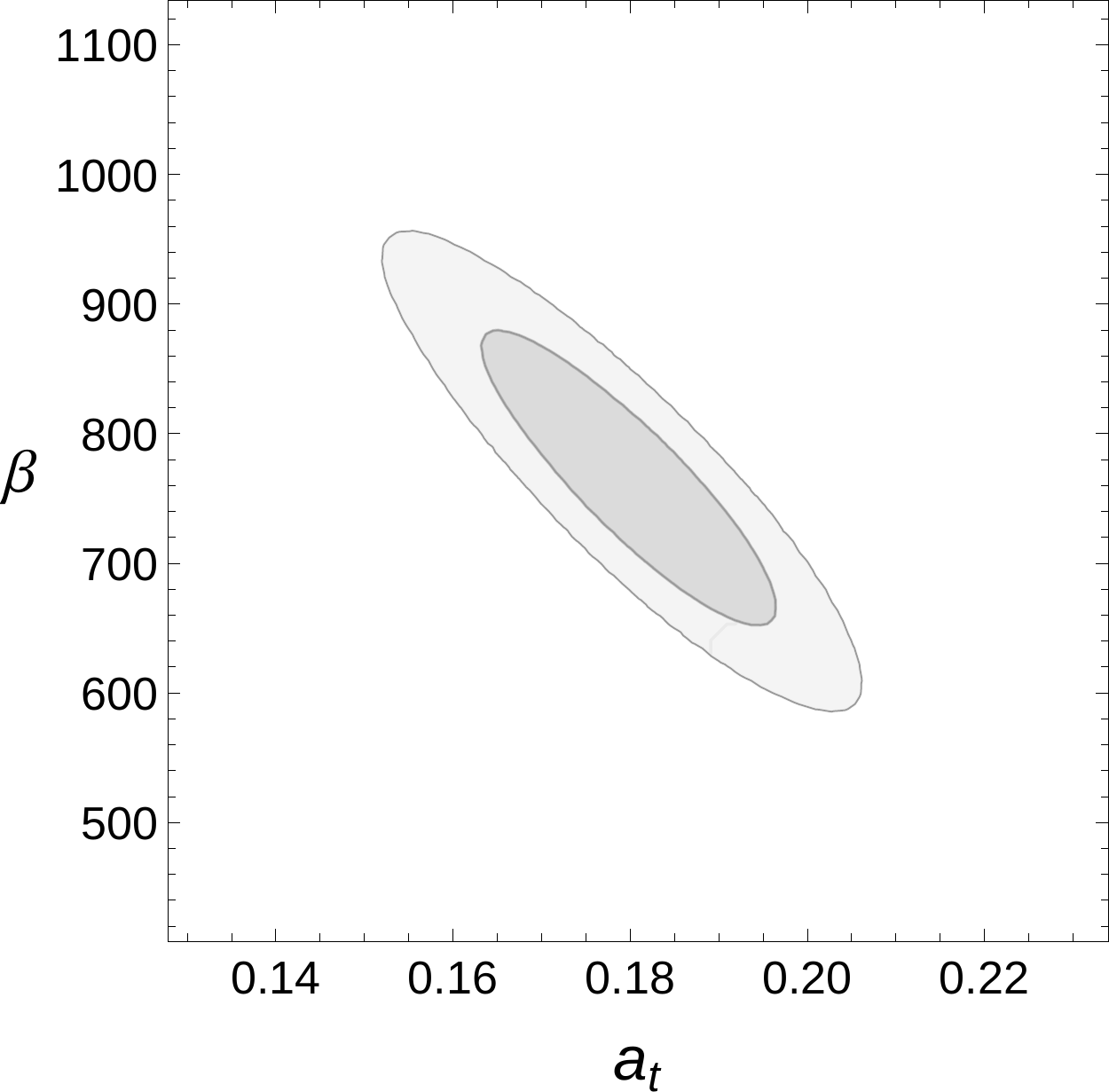}
\caption{Confidence regions for the arctan (top) and tanh (bottom) models; dark grey areas are $1\sigma$ region and light grey areas are $2\sigma$ region.} \label{fig:arctan}
\end{minipage}
\end{figure*}

\begin{figure*}[htbp]
\begin{minipage}{1.\textwidth}
 \centering
  \includegraphics[width=0.4\textwidth]{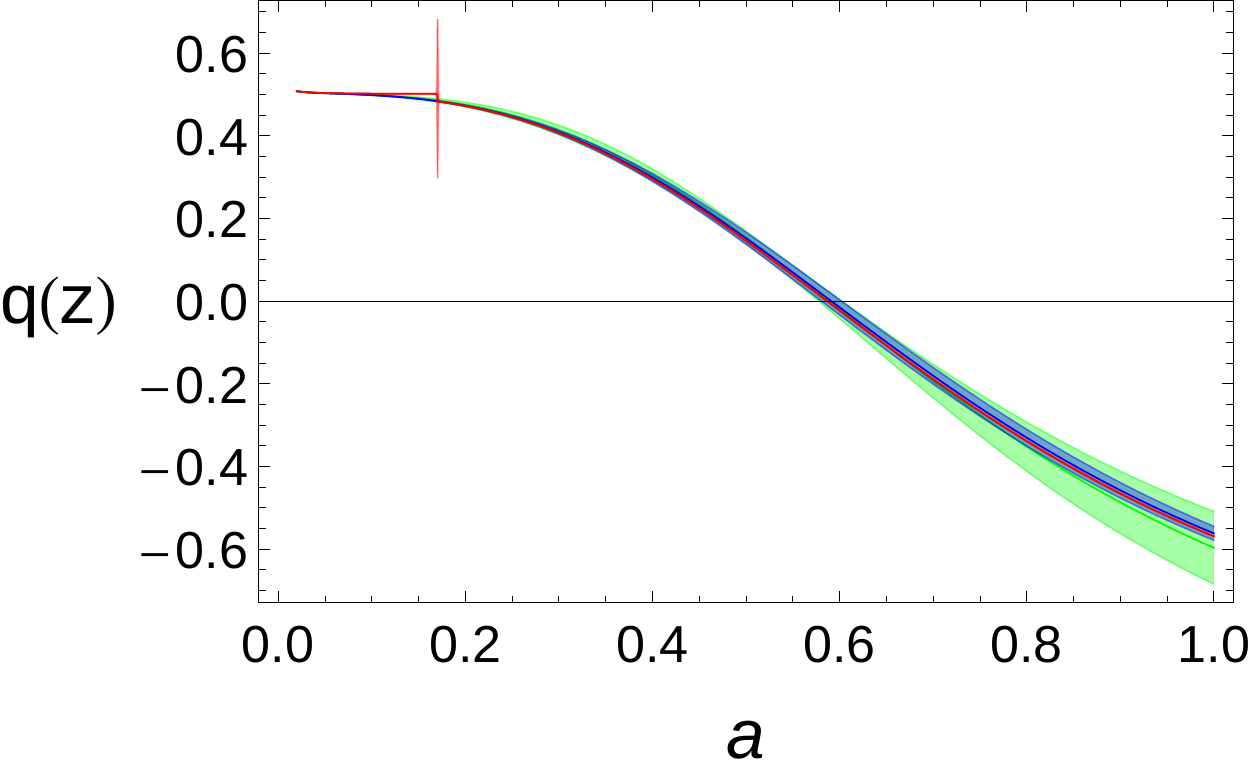}~~~~~
  \includegraphics[width=0.4\textwidth]{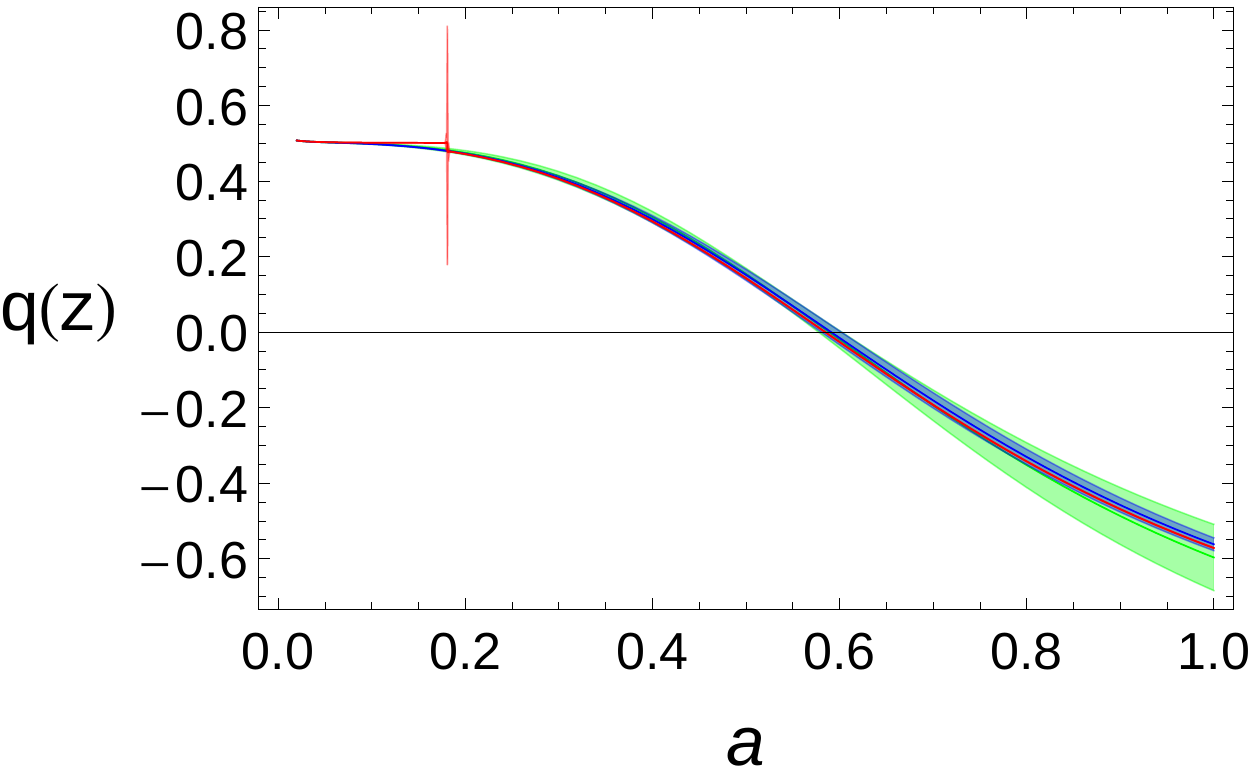}\\
  ~~~\\
  \includegraphics[width=0.4\textwidth]{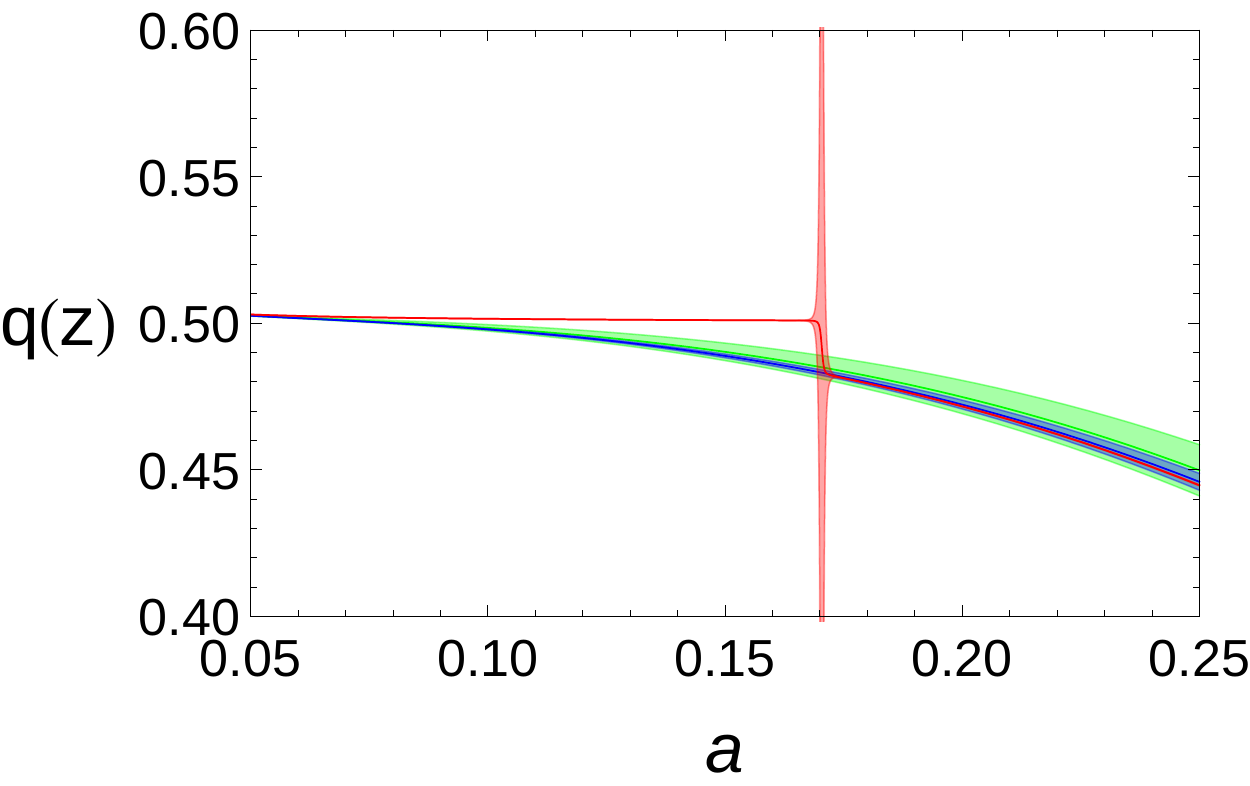}~~~~~
  \includegraphics[width=0.4\textwidth]{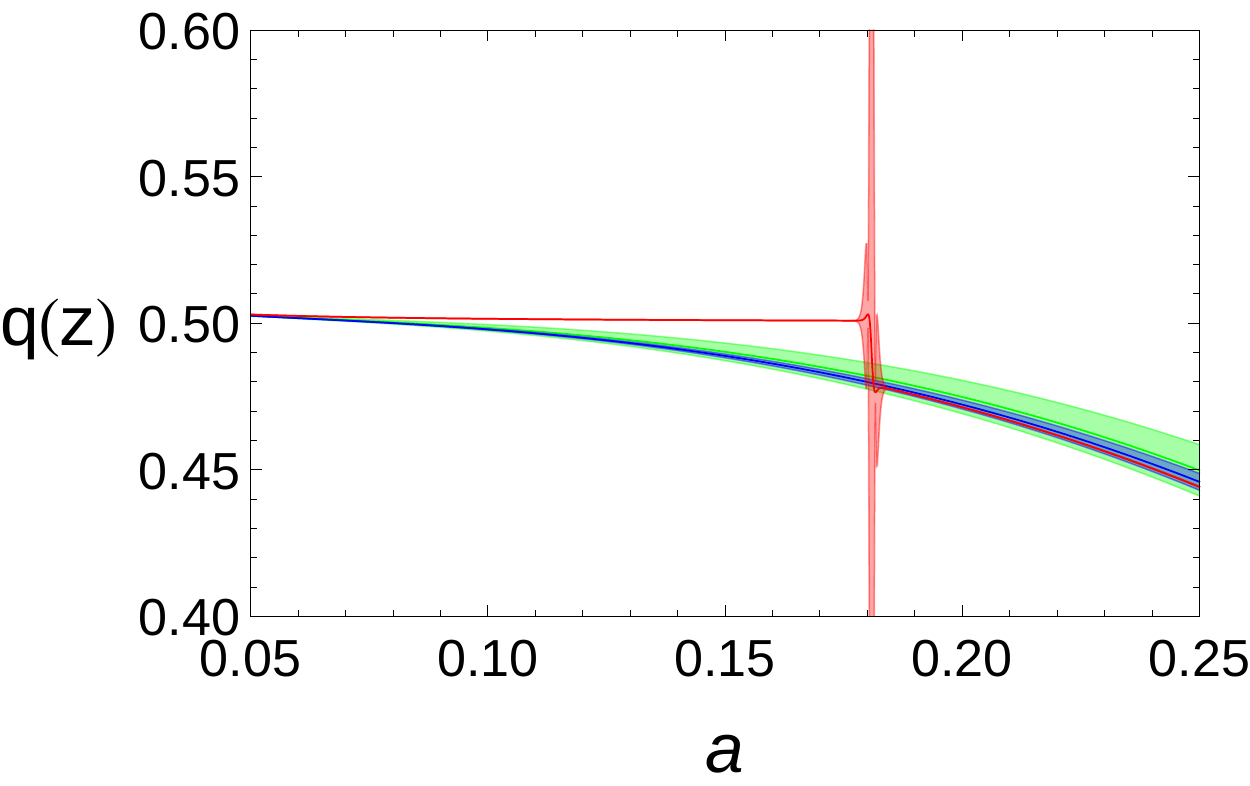}\\
\caption{Evolution of the deceleration parameter for the arctan (left, red) and tanh (right, red) models, compared with the deceleration of the $\Lambda$CDM (blue) and quiessence (green). Top figures show the entire evolution, bottom figures are zoomed in the transition.} \label{fig:q}
\end{minipage}
\end{figure*}

Fig.~\ref{fig:arctan} show the constraints on the free parameters for the models Eq.~\ref{artcan} and Eq.~\ref{tanh} respectively, and the Table~\ref{tablaSN} summarise the results.
The late time dark energy density $\Omega_{\Lambda}$ is also computed in this case, evaluating the Eq.~\ref{dark-energy} and inferring its statistics from the MCMC
output on the main fitted parameters. We also study the deceleration function, which is used as a further marker to characterize the behaviour of our models and, eventually, to better distinguish them from $\Lambda$CDM:
\be
q = -1 - \dfrac{d \log E(a)}{d \log a} .
\label{deceleration}
\ee
The deceleration function evaluated today $q_0 \equiv q(a=1)$ is also shown on the Table~\ref{tablaSN}, while its global evolution is represented in Fig.~\ref{fig:q}.

Results from the Bayesian Evidence show that the proposed UDE/M models cannot be discarded in favour of the $\Lambda$CDM model, as it can be seen in the Table~\ref{tablaSN}. All models get a very similar evidence compared each other. Even if the difference, according to the so-called ``Jeffreys' scale'' \citep{Gordon:2007xm}, still falls in the ``inconclusive evidence'' range for all of them when compared to $\Lambda$CDM, we have to highlight the higher values obtained by our UDE/M models.

When comparing the same parameters in different models, we can see that in all  four models the best fits have similar values. Nevertheless, comparing the errors of the the UDE/M models to the ones of $\Lambda$CDM and quiessence models, the former ones are slightly smaller than last ones.

These differences in the errors can be well appreciated in the plot of the deceleration \ref{fig:q}, where the differences are highlighted by the
dependence of the deceleration on all free parameters of the models. We can see that, in general,
the biggest error corresponds to the quiessence model, and the lowest the UDE/M models. However, during the transition,
due to the rapid change of the transition function, the error of the deceleration in the UDE/M models greatly increases.
We can also appreciate that the slight difference between the deceleration function of the UDE/M models
and the $\Lambda$CDM model progressively increases until the transition, but then  both UDE/M and $\Lambda$CDM models  decelerate in almost the same manner. Besides, the deceleration function of all the models,
including quiessence, has a quite similar behaviour (and value range) during the whole evolution of the universe.

This transition is characterised by the transition functions \ref{artcan} and \ref{tanh},
which is shaped by the parameters $a_t$ and $\beta$. For both UDE/M models the transition occurs in the past,
and is centered around $a_t \approxeq 0.17-0.18$. The transitions occur very fast, in a narrow fraction of the entire cosmic time.

The mayor differences between the UDE/M models and $\Lambda$CDM occur before the transition.
In this way, they could be considered as early time deviations from $\Lambda$CDM, where there would be no need for dark energy in the past.
Furthermore, the dark sector of the universe could be explained by a single UDE/M fluid instead of the two components necessary with $\Lambda$CDM.\\

\section{Acknowledgements}
We are grateful to Marco Bruni for enlightening and insightful conversations. We also thank Alberto Rozas Fern\'andez for useful comments. 
R.L. and I.L. were supported by the Spanish Ministry of Economy and Competitiveness through research projects FIS2014-57956-P (comprising FEDER funds) and also by the Basque Government through research project GIC12/66, and by the University of the Basque Country UPV/EHU under program UFI 11/55. 
R.L. and V.S were also supported by the above last mentioned ministry through  research grant Consolider EPI CSD2010-00064.
I.L. acknowledges financial support from the University of the Basque Country UPV/EHU PhD grant 750/2014.
V.S. is funded by the Polish National Science Center Grant DEC-2012/06/A/ST2/00395.

\bibliography{draft-referee}

\begin{thebibliography}{52}%
\makeatletter
\providecommand \@ifxundefined [1]{%
 \@ifx{#1\undefined}
}%
\providecommand \@ifnum [1]{%
 \ifnum #1\expandafter \@firstoftwo
 \else \expandafter \@secondoftwo
 \fi
}%
\providecommand \@ifx [1]{%
 \ifx #1\expandafter \@firstoftwo
 \else \expandafter \@secondoftwo
 \fi
}%
\providecommand \natexlab [1]{#1}%
\providecommand \enquote  [1]{``#1''}%
\providecommand \bibnamefont  [1]{#1}%
\providecommand \bibfnamefont [1]{#1}%
\providecommand \citenamefont [1]{#1}%
\providecommand \href@noop [0]{\@secondoftwo}%
\providecommand \href [0]{\begingroup \@sanitize@url \@href}%
\providecommand \@href[1]{\@@startlink{#1}\@@href}%
\providecommand \@@href[1]{\endgroup#1\@@endlink}%
\providecommand \@sanitize@url [0]{\catcode `\\12\catcode `\$12\catcode
  `\&12\catcode `\#12\catcode `\^12\catcode `\_12\catcode `\%12\relax}%
\providecommand \@@startlink[1]{}%
\providecommand \@@endlink[0]{}%
\providecommand \url  [0]{\begingroup\@sanitize@url \@url }%
\providecommand \@url [1]{\endgroup\@href {#1}{\urlprefix }}%
\providecommand \urlprefix  [0]{URL }%
\providecommand \Eprint [0]{\href }%
\providecommand \doibase [0]{http://dx.doi.org/}%
\providecommand \selectlanguage [0]{\@gobble}%
\providecommand \bibinfo  [0]{\@secondoftwo}%
\providecommand \bibfield  [0]{\@secondoftwo}%
\providecommand \translation [1]{[#1]}%
\providecommand \BibitemOpen [0]{}%
\providecommand \bibitemStop [0]{}%
\providecommand \bibitemNoStop [0]{.\EOS\space}%
\providecommand \EOS [0]{\spacefactor3000\relax}%
\providecommand \BibitemShut  [1]{\csname bibitem#1\endcsname}%
\let\auto@bib@innerbib\@empty
\bibitem [{\citenamefont {Riess}\ \emph {et~al.}(1998)\citenamefont {Riess}
  \emph {et~al.}}]{Riess:1998cb}%
  \BibitemOpen
  \bibfield  {author} {\bibinfo {author} {\bibfnamefont {A.~G.}\ \bibnamefont
  {Riess}} \emph {et~al.} (\bibinfo {collaboration} {Supernova Search Team}),\
  }\href {\doibase 10.1086/300499} {\bibfield  {journal} {\bibinfo  {journal}
  {Astron.J.}\ }\textbf {\bibinfo {volume} {116}},\ \bibinfo {pages} {1009}
  (\bibinfo {year} {1998})},\ \Eprint {http://arxiv.org/abs/astro-ph/9805201}
  {arXiv:astro-ph/9805201 [astro-ph]} \BibitemShut {NoStop}%
\bibitem [{\citenamefont {Perlmutter}\ \emph {et~al.}(1999)\citenamefont
  {Perlmutter} \emph {et~al.}}]{Perlmutter:1998np}%
  \BibitemOpen
  \bibfield  {author} {\bibinfo {author} {\bibfnamefont {S.}~\bibnamefont
  {Perlmutter}} \emph {et~al.} (\bibinfo {collaboration} {Supernova Cosmology
  Project}),\ }\href {\doibase 10.1086/307221} {\bibfield  {journal} {\bibinfo
  {journal} {Astrophys.J.}\ }\textbf {\bibinfo {volume} {517}},\ \bibinfo
  {pages} {565} (\bibinfo {year} {1999})},\ \Eprint
  {http://arxiv.org/abs/astro-ph/9812133} {arXiv:astro-ph/9812133 [astro-ph]}
  \BibitemShut {NoStop}%
\bibitem [{\citenamefont {Ade}\ \emph {et~al.}(2014)\citenamefont {Ade} \emph
  {et~al.}}]{Ade:2013zuv}%
  \BibitemOpen
  \bibfield  {author} {\bibinfo {author} {\bibfnamefont {P.}~\bibnamefont
  {Ade}} \emph {et~al.} (\bibinfo {collaboration} {Planck Collaboration}),\
  }\href {\doibase 10.1051/0004-6361/201321591} {\bibfield  {journal} {\bibinfo
   {journal} {Astron.Astrophys.}\ }\textbf {\bibinfo {volume} {571}},\ \bibinfo
  {pages} {A16} (\bibinfo {year} {2014})},\ \Eprint
  {http://arxiv.org/abs/1303.5076} {arXiv:1303.5076 [astro-ph.CO]} \BibitemShut
  {NoStop}%
\bibitem [{\citenamefont {Weinberg}\ \emph {et~al.}(2013)\citenamefont
  {Weinberg}, \citenamefont {Mortonson}, \citenamefont {Eisenstein},
  \citenamefont {Hirata}, \citenamefont {Riess} \emph
  {et~al.}}]{Weinberg:2012es}%
  \BibitemOpen
  \bibfield  {author} {\bibinfo {author} {\bibfnamefont {D.~H.}\ \bibnamefont
  {Weinberg}}, \bibinfo {author} {\bibfnamefont {M.~J.}\ \bibnamefont
  {Mortonson}}, \bibinfo {author} {\bibfnamefont {D.~J.}\ \bibnamefont
  {Eisenstein}}, \bibinfo {author} {\bibfnamefont {C.}~\bibnamefont {Hirata}},
  \bibinfo {author} {\bibfnamefont {A.~G.}\ \bibnamefont {Riess}},  \emph
  {et~al.},\ }\href {\doibase 10.1016/j.physrep.2013.05.001} {\bibfield
  {journal} {\bibinfo  {journal} {Phys.Rept.}\ }\textbf {\bibinfo {volume}
  {530}},\ \bibinfo {pages} {87} (\bibinfo {year} {2013})},\ \Eprint
  {http://arxiv.org/abs/1201.2434} {arXiv:1201.2434 [astro-ph.CO]} \BibitemShut
  {NoStop}%
\bibitem [{\citenamefont {Carroll}(2001)}]{Carroll:2000fy}%
  \BibitemOpen
  \bibfield  {author} {\bibinfo {author} {\bibfnamefont {S.~M.}\ \bibnamefont
  {Carroll}},\ }\href@noop {} {\bibfield  {journal} {\bibinfo  {journal}
  {Living Rev.Rel.}\ }\textbf {\bibinfo {volume} {4}},\ \bibinfo {pages} {1}
  (\bibinfo {year} {2001})},\ \Eprint {http://arxiv.org/abs/astro-ph/0004075}
  {arXiv:astro-ph/0004075 [astro-ph]} \BibitemShut {NoStop}%
\bibitem [{\citenamefont {Berti}\ \emph {et~al.}(2015)\citenamefont {Berti},
  \citenamefont {Barausse}, \citenamefont {Cardoso}, \citenamefont {Gualtieri},
  \citenamefont {Pani} \emph {et~al.}}]{Berti:2015aea}%
  \BibitemOpen
  \bibfield  {author} {\bibinfo {author} {\bibfnamefont {E.}~\bibnamefont
  {Berti}}, \bibinfo {author} {\bibfnamefont {E.}~\bibnamefont {Barausse}},
  \bibinfo {author} {\bibfnamefont {V.}~\bibnamefont {Cardoso}}, \bibinfo
  {author} {\bibfnamefont {L.}~\bibnamefont {Gualtieri}}, \bibinfo {author}
  {\bibfnamefont {P.}~\bibnamefont {Pani}},  \emph {et~al.},\ }\href@noop {} {\
   (\bibinfo {year} {2015})},\ \Eprint {http://arxiv.org/abs/1501.07274}
  {arXiv:1501.07274 [gr-qc]} \BibitemShut {NoStop}%
\bibitem [{\citenamefont {Li}\ \emph {et~al.}(2013)\citenamefont {Li},
  \citenamefont {Li}, \citenamefont {Wang},\ and\ \citenamefont
  {Wang}}]{Li:2012dt}%
  \BibitemOpen
  \bibfield  {author} {\bibinfo {author} {\bibfnamefont {M.}~\bibnamefont
  {Li}}, \bibinfo {author} {\bibfnamefont {X.-D.}\ \bibnamefont {Li}}, \bibinfo
  {author} {\bibfnamefont {S.}~\bibnamefont {Wang}}, \ and\ \bibinfo {author}
  {\bibfnamefont {Y.}~\bibnamefont {Wang}},\ }\href {\doibase
  10.1007/s11467-013-0300-5} {\bibfield  {journal} {\bibinfo  {journal}
  {Front.Phys.China}\ }\textbf {\bibinfo {volume} {8}},\ \bibinfo {pages} {828}
  (\bibinfo {year} {2013})},\ \Eprint {http://arxiv.org/abs/1209.0922}
  {arXiv:1209.0922 [astro-ph.CO]} \BibitemShut {NoStop}%
\bibitem [{\citenamefont {Kunz}(2012)}]{Kunz:2012aw}%
  \BibitemOpen
  \bibfield  {author} {\bibinfo {author} {\bibfnamefont {M.}~\bibnamefont
  {Kunz}},\ }\href {\doibase 10.1016/j.crhy.2012.04.007} {\bibfield  {journal}
  {\bibinfo  {journal} {Comptes Rendus Physique}\ }\textbf {\bibinfo {volume}
  {13}},\ \bibinfo {pages} {539} (\bibinfo {year} {2012})},\ \Eprint
  {http://arxiv.org/abs/1204.5482} {arXiv:1204.5482 [astro-ph.CO]} \BibitemShut
  {NoStop}%
\bibitem [{\citenamefont {Copeland}\ \emph {et~al.}(2006)\citenamefont
  {Copeland}, \citenamefont {Sami},\ and\ \citenamefont
  {Tsujikawa}}]{Copeland:2006wr}%
  \BibitemOpen
  \bibfield  {author} {\bibinfo {author} {\bibfnamefont {E.~J.}\ \bibnamefont
  {Copeland}}, \bibinfo {author} {\bibfnamefont {M.}~\bibnamefont {Sami}}, \
  and\ \bibinfo {author} {\bibfnamefont {S.}~\bibnamefont {Tsujikawa}},\ }\href
  {\doibase 10.1142/S021827180600942X} {\bibfield  {journal} {\bibinfo
  {journal} {Int.J.Mod.Phys.}\ }\textbf {\bibinfo {volume} {D15}},\ \bibinfo
  {pages} {1753} (\bibinfo {year} {2006})},\ \Eprint
  {http://arxiv.org/abs/hep-th/0603057} {arXiv:hep-th/0603057 [hep-th]}
  \BibitemShut {NoStop}%
\bibitem [{\citenamefont {Bamba}\ \emph {et~al.}(2012)\citenamefont {Bamba},
  \citenamefont {Capozziello}, \citenamefont {Nojiri},\ and\ \citenamefont
  {Odintsov}}]{Bamba:2012cp}%
  \BibitemOpen
  \bibfield  {author} {\bibinfo {author} {\bibfnamefont {K.}~\bibnamefont
  {Bamba}}, \bibinfo {author} {\bibfnamefont {S.}~\bibnamefont {Capozziello}},
  \bibinfo {author} {\bibfnamefont {S.}~\bibnamefont {Nojiri}}, \ and\ \bibinfo
  {author} {\bibfnamefont {S.~D.}\ \bibnamefont {Odintsov}},\ }\href {\doibase
  10.1007/s10509-012-1181-8} {\bibfield  {journal} {\bibinfo  {journal}
  {Astrophys.Space Sci.}\ }\textbf {\bibinfo {volume} {342}},\ \bibinfo {pages}
  {155} (\bibinfo {year} {2012})},\ \Eprint {http://arxiv.org/abs/1205.3421}
  {arXiv:1205.3421 [gr-qc]} \BibitemShut {NoStop}%
\bibitem [{\citenamefont {Zwicky}(1933)}]{Zwicky:1933gu}%
  \BibitemOpen
  \bibfield  {author} {\bibinfo {author} {\bibfnamefont {F.}~\bibnamefont
  {Zwicky}},\ }\href@noop {} {\bibfield  {journal} {\bibinfo  {journal}
  {Helv.Phys.Acta}\ }\textbf {\bibinfo {volume} {6}},\ \bibinfo {pages} {110}
  (\bibinfo {year} {1933})}\BibitemShut {NoStop}%
\bibitem [{\citenamefont {Rubin}\ \emph {et~al.}(1980)\citenamefont {Rubin},
  \citenamefont {Thonnard},\ and\ \citenamefont {Ford}}]{Rubin:1980zd}%
  \BibitemOpen
  \bibfield  {author} {\bibinfo {author} {\bibfnamefont {V.}~\bibnamefont
  {Rubin}}, \bibinfo {author} {\bibfnamefont {N.}~\bibnamefont {Thonnard}}, \
  and\ \bibinfo {author} {\bibfnamefont {J.}~\bibnamefont {Ford}, \bibfnamefont
  {W.K.}},\ }\href {\doibase 10.1086/158003} {\bibfield  {journal} {\bibinfo
  {journal} {Astrophys.J.}\ }\textbf {\bibinfo {volume} {238}},\ \bibinfo
  {pages} {471} (\bibinfo {year} {1980})}\BibitemShut {NoStop}%
\bibitem [{\citenamefont {Peter}(2011)}]{Peter:2012rz}%
  \BibitemOpen
  \bibfield  {author} {\bibinfo {author} {\bibfnamefont {A.~H.}\ \bibnamefont
  {Peter}},\ }\href@noop {} {\bibfield  {journal} {\bibinfo  {journal}
  {{Proceedings of Science (Bash11)}}\ }\textbf {\bibinfo {volume} {014}}
  (\bibinfo {year} {2011})},\ \Eprint {http://arxiv.org/abs/1201.3942}
  {arXiv:1201.3942 [astro-ph.CO]} \BibitemShut {NoStop}%
\bibitem [{\citenamefont {Lukovic}\ \emph {et~al.}(2014)\citenamefont
  {Lukovic}, \citenamefont {Cabella},\ and\ \citenamefont
  {Vittorio}}]{Lukovic:2014vma}%
  \BibitemOpen
  \bibfield  {author} {\bibinfo {author} {\bibfnamefont {V.}~\bibnamefont
  {Lukovic}}, \bibinfo {author} {\bibfnamefont {P.}~\bibnamefont {Cabella}}, \
  and\ \bibinfo {author} {\bibfnamefont {N.}~\bibnamefont {Vittorio}},\ }\href
  {\doibase 10.1142/S0217751X14430015} {\bibfield  {journal} {\bibinfo
  {journal} {Int.J.Mod.Phys.}\ }\textbf {\bibinfo {volume} {A29}},\ \bibinfo
  {pages} {1443001} (\bibinfo {year} {2014})},\ \Eprint
  {http://arxiv.org/abs/1411.3556} {arXiv:1411.3556 [astro-ph.CO]} \BibitemShut
  {NoStop}%
\bibitem [{\citenamefont {Einasto}(2009)}]{Einasto:2009zd}%
  \BibitemOpen
  \bibfield  {author} {\bibinfo {author} {\bibfnamefont {J.}~\bibnamefont
  {Einasto}},\ }\href@noop {} {\  (\bibinfo {year} {2009})},\ \Eprint
  {http://arxiv.org/abs/0901.0632} {arXiv:0901.0632 [astro-ph.CO]} \BibitemShut
  {NoStop}%
\bibitem [{\citenamefont {Capozziello}\ \emph {et~al.}(2011)\citenamefont
  {Capozziello}, \citenamefont {Consiglio}, \citenamefont {De~Laurentis},
  \citenamefont {De~Rosa},\ and\ \citenamefont
  {Di~Donato}}]{Capozziello:2011xp}%
  \BibitemOpen
  \bibfield  {author} {\bibinfo {author} {\bibfnamefont {S.}~\bibnamefont
  {Capozziello}}, \bibinfo {author} {\bibfnamefont {L.}~\bibnamefont
  {Consiglio}}, \bibinfo {author} {\bibfnamefont {M.}~\bibnamefont
  {De~Laurentis}}, \bibinfo {author} {\bibfnamefont {G.}~\bibnamefont
  {De~Rosa}}, \ and\ \bibinfo {author} {\bibfnamefont {C.}~\bibnamefont
  {Di~Donato}},\ }\href@noop {} {\  (\bibinfo {year} {2011})},\ \Eprint
  {http://arxiv.org/abs/1110.5026} {arXiv:1110.5026 [astro-ph.CO]} \BibitemShut
  {NoStop}%
\bibitem [{\citenamefont {Kamenshchik}\ \emph {et~al.}(2001)\citenamefont
  {Kamenshchik}, \citenamefont {Moschella},\ and\ \citenamefont
  {Pasquier}}]{Kamenshchik:2001cp}%
  \BibitemOpen
  \bibfield  {author} {\bibinfo {author} {\bibfnamefont {A.~Y.}\ \bibnamefont
  {Kamenshchik}}, \bibinfo {author} {\bibfnamefont {U.}~\bibnamefont
  {Moschella}}, \ and\ \bibinfo {author} {\bibfnamefont {V.}~\bibnamefont
  {Pasquier}},\ }\href {\doibase 10.1016/S0370-2693(01)00571-8} {\bibfield
  {journal} {\bibinfo  {journal} {Phys.Lett.}\ }\textbf {\bibinfo {volume}
  {B511}},\ \bibinfo {pages} {265} (\bibinfo {year} {2001})},\ \Eprint
  {http://arxiv.org/abs/gr-qc/0103004} {arXiv:gr-qc/0103004 [gr-qc]}
  \BibitemShut {NoStop}%
\bibitem [{\citenamefont {Bilic}\ \emph {et~al.}(2002)\citenamefont {Bilic},
  \citenamefont {Tupper},\ and\ \citenamefont {Viollier}}]{Bilic:2001cg}%
  \BibitemOpen
  \bibfield  {author} {\bibinfo {author} {\bibfnamefont {N.}~\bibnamefont
  {Bilic}}, \bibinfo {author} {\bibfnamefont {G.~B.}\ \bibnamefont {Tupper}}, \
  and\ \bibinfo {author} {\bibfnamefont {R.~D.}\ \bibnamefont {Viollier}},\
  }\href {\doibase 10.1016/S0370-2693(02)01716-1} {\bibfield  {journal}
  {\bibinfo  {journal} {Phys.Lett.}\ }\textbf {\bibinfo {volume} {B535}},\
  \bibinfo {pages} {17} (\bibinfo {year} {2002})},\ \Eprint
  {http://arxiv.org/abs/astro-ph/0111325} {arXiv:astro-ph/0111325 [astro-ph]}
  \BibitemShut {NoStop}%
\bibitem [{\citenamefont {Bertacca}\ \emph {et~al.}(2010)\citenamefont
  {Bertacca}, \citenamefont {Bartolo},\ and\ \citenamefont
  {Matarrese}}]{Bertacca:2010ct}%
  \BibitemOpen
  \bibfield  {author} {\bibinfo {author} {\bibfnamefont {D.}~\bibnamefont
  {Bertacca}}, \bibinfo {author} {\bibfnamefont {N.}~\bibnamefont {Bartolo}}, \
  and\ \bibinfo {author} {\bibfnamefont {S.}~\bibnamefont {Matarrese}},\ }\href
  {\doibase 10.1155/2010/904379} {\bibfield  {journal} {\bibinfo  {journal}
  {Adv.Astron.}\ }\textbf {\bibinfo {volume} {2010}},\ \bibinfo {pages}
  {904379} (\bibinfo {year} {2010})},\ \Eprint {http://arxiv.org/abs/1008.0614}
  {arXiv:1008.0614 [astro-ph.CO]} \BibitemShut {NoStop}%
\bibitem [{\citenamefont {Amendola}(2000)}]{Amendola:1999er}%
  \BibitemOpen
  \bibfield  {author} {\bibinfo {author} {\bibfnamefont {L.}~\bibnamefont
  {Amendola}},\ }\href {\doibase 10.1103/PhysRevD.62.043511} {\bibfield
  {journal} {\bibinfo  {journal} {Phys. Rev.}\ }\textbf {\bibinfo {volume}
  {D62}},\ \bibinfo {pages} {043511} (\bibinfo {year} {2000})},\ \Eprint
  {http://arxiv.org/abs/astro-ph/9908023} {arXiv:astro-ph/9908023 [astro-ph]}
  \BibitemShut {NoStop}%
\bibitem [{\citenamefont {Salvatelli}\ \emph {et~al.}(2014)\citenamefont
  {Salvatelli}, \citenamefont {Said}, \citenamefont {Bruni}, \citenamefont
  {Melchiorri},\ and\ \citenamefont {Wands}}]{Salvatelli:2014zta}%
  \BibitemOpen
  \bibfield  {author} {\bibinfo {author} {\bibfnamefont {V.}~\bibnamefont
  {Salvatelli}}, \bibinfo {author} {\bibfnamefont {N.}~\bibnamefont {Said}},
  \bibinfo {author} {\bibfnamefont {M.}~\bibnamefont {Bruni}}, \bibinfo
  {author} {\bibfnamefont {A.}~\bibnamefont {Melchiorri}}, \ and\ \bibinfo
  {author} {\bibfnamefont {D.}~\bibnamefont {Wands}},\ }\href {\doibase
  10.1103/PhysRevLett.113.181301} {\bibfield  {journal} {\bibinfo  {journal}
  {Phys. Rev. Lett.}\ }\textbf {\bibinfo {volume} {113}},\ \bibinfo {pages}
  {181301} (\bibinfo {year} {2014})},\ \Eprint {http://arxiv.org/abs/1406.7297}
  {arXiv:1406.7297 [astro-ph.CO]} \BibitemShut {NoStop}%
\bibitem [{\citenamefont {Valiviita}\ and\ \citenamefont
  {Palmgren}(2015)}]{Valiviita:2015dfa}%
  \BibitemOpen
  \bibfield  {author} {\bibinfo {author} {\bibfnamefont {J.}~\bibnamefont
  {Valiviita}}\ and\ \bibinfo {author} {\bibfnamefont {E.}~\bibnamefont
  {Palmgren}},\ }\href {\doibase 10.1088/1475-7516/2015/07/015} {\bibfield
  {journal} {\bibinfo  {journal} {JCAP}\ }\textbf {\bibinfo {volume} {1507}},\
  \bibinfo {pages} {015} (\bibinfo {year} {2015})},\ \Eprint
  {http://arxiv.org/abs/1504.02464} {arXiv:1504.02464 [astro-ph.CO]}
  \BibitemShut {NoStop}%
\bibitem [{\citenamefont {Sandvik}\ \emph {et~al.}(2004)\citenamefont
  {Sandvik}, \citenamefont {Tegmark}, \citenamefont {Zaldarriaga},\ and\
  \citenamefont {Waga}}]{Sandvik:2002jz}%
  \BibitemOpen
  \bibfield  {author} {\bibinfo {author} {\bibfnamefont {H.}~\bibnamefont
  {Sandvik}}, \bibinfo {author} {\bibfnamefont {M.}~\bibnamefont {Tegmark}},
  \bibinfo {author} {\bibfnamefont {M.}~\bibnamefont {Zaldarriaga}}, \ and\
  \bibinfo {author} {\bibfnamefont {I.}~\bibnamefont {Waga}},\ }\href {\doibase
  10.1103/PhysRevD.69.123524} {\bibfield  {journal} {\bibinfo  {journal}
  {Phys.Rev.}\ }\textbf {\bibinfo {volume} {D69}},\ \bibinfo {pages} {123524}
  (\bibinfo {year} {2004})},\ \Eprint {http://arxiv.org/abs/astro-ph/0212114}
  {arXiv:astro-ph/0212114 [astro-ph]} \BibitemShut {NoStop}%
\bibitem [{\citenamefont {Bilic}\ \emph {et~al.}(2004)\citenamefont {Bilic},
  \citenamefont {Lindebaum}, \citenamefont {Tupper},\ and\ \citenamefont
  {Viollier}}]{Bilic:2003cv}%
  \BibitemOpen
  \bibfield  {author} {\bibinfo {author} {\bibfnamefont {N.}~\bibnamefont
  {Bilic}}, \bibinfo {author} {\bibfnamefont {R.~J.}\ \bibnamefont
  {Lindebaum}}, \bibinfo {author} {\bibfnamefont {G.~B.}\ \bibnamefont
  {Tupper}}, \ and\ \bibinfo {author} {\bibfnamefont {R.~D.}\ \bibnamefont
  {Viollier}},\ }\href {\doibase 10.1088/1475-7516/2004/11/008} {\bibfield
  {journal} {\bibinfo  {journal} {JCAP}\ }\textbf {\bibinfo {volume} {0411}},\
  \bibinfo {pages} {008} (\bibinfo {year} {2004})},\ \Eprint
  {http://arxiv.org/abs/astro-ph/0307214} {arXiv:astro-ph/0307214 [astro-ph]}
  \BibitemShut {NoStop}%
\bibitem [{\citenamefont {Piattella}\ \emph {et~al.}(2010)\citenamefont
  {Piattella}, \citenamefont {Bertacca}, \citenamefont {Bruni},\ and\
  \citenamefont {Pietrobon}}]{Piattella:2009kt}%
  \BibitemOpen
  \bibfield  {author} {\bibinfo {author} {\bibfnamefont {O.~F.}\ \bibnamefont
  {Piattella}}, \bibinfo {author} {\bibfnamefont {D.}~\bibnamefont {Bertacca}},
  \bibinfo {author} {\bibfnamefont {M.}~\bibnamefont {Bruni}}, \ and\ \bibinfo
  {author} {\bibfnamefont {D.}~\bibnamefont {Pietrobon}},\ }\href {\doibase
  10.1088/1475-7516/2010/01/014} {\bibfield  {journal} {\bibinfo  {journal}
  {JCAP}\ }\textbf {\bibinfo {volume} {1001}},\ \bibinfo {pages} {014}
  (\bibinfo {year} {2010})},\ \Eprint {http://arxiv.org/abs/0911.2664}
  {arXiv:0911.2664 [astro-ph.CO]} \BibitemShut {NoStop}%
\bibitem [{\citenamefont {Bertacca}\ \emph {et~al.}(2011)\citenamefont
  {Bertacca}, \citenamefont {Bruni}, \citenamefont {Piattella},\ and\
  \citenamefont {Pietrobon}}]{Bertacca:2010mt}%
  \BibitemOpen
  \bibfield  {author} {\bibinfo {author} {\bibfnamefont {D.}~\bibnamefont
  {Bertacca}}, \bibinfo {author} {\bibfnamefont {M.}~\bibnamefont {Bruni}},
  \bibinfo {author} {\bibfnamefont {O.~F.}\ \bibnamefont {Piattella}}, \ and\
  \bibinfo {author} {\bibfnamefont {D.}~\bibnamefont {Pietrobon}},\ }\href
  {\doibase 10.1088/1475-7516/2011/02/018} {\bibfield  {journal} {\bibinfo
  {journal} {JCAP}\ }\textbf {\bibinfo {volume} {1102}},\ \bibinfo {pages}
  {018} (\bibinfo {year} {2011})},\ \Eprint {http://arxiv.org/abs/1011.6669}
  {arXiv:1011.6669 [astro-ph.CO]} \BibitemShut {NoStop}%
\bibitem [{\citenamefont {Bruni}\ \emph {et~al.}(2013)\citenamefont {Bruni},
  \citenamefont {Lazkoz},\ and\ \citenamefont
  {Rozas-Fernandez}}]{Bruni:2012sn}%
  \BibitemOpen
  \bibfield  {author} {\bibinfo {author} {\bibfnamefont {M.}~\bibnamefont
  {Bruni}}, \bibinfo {author} {\bibfnamefont {R.}~\bibnamefont {Lazkoz}}, \
  and\ \bibinfo {author} {\bibfnamefont {A.}~\bibnamefont {Rozas-Fernandez}},\
  }\href {\doibase 10.1093/mnras/stt395} {\bibfield  {journal} {\bibinfo
  {journal} {Mon.Not.Roy.Astron.Soc.}\ }\textbf {\bibinfo {volume} {431}},\
  \bibinfo {pages} {2907} (\bibinfo {year} {2013})},\ \Eprint
  {http://arxiv.org/abs/1210.1880} {arXiv:1210.1880 [astro-ph.CO]} \BibitemShut
  {NoStop}%
\bibitem [{\citenamefont {Wang}\ \emph {et~al.}(2013)\citenamefont {Wang},
  \citenamefont {Wands}, \citenamefont {Xu}, \citenamefont {De-Santiago},\ and\
  \citenamefont {Hojjati}}]{Wang:2013qy}%
  \BibitemOpen
  \bibfield  {author} {\bibinfo {author} {\bibfnamefont {Y.}~\bibnamefont
  {Wang}}, \bibinfo {author} {\bibfnamefont {D.}~\bibnamefont {Wands}},
  \bibinfo {author} {\bibfnamefont {L.}~\bibnamefont {Xu}}, \bibinfo {author}
  {\bibfnamefont {J.}~\bibnamefont {De-Santiago}}, \ and\ \bibinfo {author}
  {\bibfnamefont {A.}~\bibnamefont {Hojjati}},\ }\href {\doibase
  10.1103/PhysRevD.87.083503} {\bibfield  {journal} {\bibinfo  {journal} {Phys.
  Rev.}\ }\textbf {\bibinfo {volume} {D87}},\ \bibinfo {pages} {083503}
  (\bibinfo {year} {2013})},\ \Eprint {http://arxiv.org/abs/1301.5315}
  {arXiv:1301.5315 [astro-ph.CO]} \BibitemShut {NoStop}%
\bibitem [{\citenamefont {De-Santiago}\ \emph {et~al.}(2012)\citenamefont
  {De-Santiago}, \citenamefont {Wands},\ and\ \citenamefont
  {Wang}}]{DeSantiago:2012xh}%
  \BibitemOpen
  \bibfield  {author} {\bibinfo {author} {\bibfnamefont {J.}~\bibnamefont
  {De-Santiago}}, \bibinfo {author} {\bibfnamefont {D.}~\bibnamefont {Wands}},
  \ and\ \bibinfo {author} {\bibfnamefont {Y.}~\bibnamefont {Wang}},\ }in\
  \href@noop {} {\emph {\bibinfo {booktitle} {{6th International Meeting on
  Gravitation and Cosmology Guadalajara, Jalisco, Mexico, May 21-25, 2012}}}}\
  (\bibinfo {year} {2012})\ \Eprint {http://arxiv.org/abs/1209.0563}
  {arXiv:1209.0563 [astro-ph.CO]} \BibitemShut {NoStop}%
\bibitem [{\citenamefont {Wang}\ \emph {et~al.}(2015)\citenamefont {Wang},
  \citenamefont {Zhao}, \citenamefont {Wands}, \citenamefont {Pogosian},\ and\
  \citenamefont {Crittenden}}]{Wang:2015wga}%
  \BibitemOpen
  \bibfield  {author} {\bibinfo {author} {\bibfnamefont {Y.}~\bibnamefont
  {Wang}}, \bibinfo {author} {\bibfnamefont {G.-B.}\ \bibnamefont {Zhao}},
  \bibinfo {author} {\bibfnamefont {D.}~\bibnamefont {Wands}}, \bibinfo
  {author} {\bibfnamefont {L.}~\bibnamefont {Pogosian}}, \ and\ \bibinfo
  {author} {\bibfnamefont {R.~G.}\ \bibnamefont {Crittenden}},\ }\href
  {\doibase 10.1103/PhysRevD.92.103005} {\bibfield  {journal} {\bibinfo
  {journal} {Phys. Rev.}\ }\textbf {\bibinfo {volume} {D92}},\ \bibinfo {pages}
  {103005} (\bibinfo {year} {2015})},\ \Eprint
  {http://arxiv.org/abs/1505.01373} {arXiv:1505.01373 [astro-ph.CO]}
  \BibitemShut {NoStop}%
\bibitem [{\citenamefont {Wang}\ and\ \citenamefont {Wang}(2013)}]{Wang2013}%
  \BibitemOpen
  \bibfield  {author} {\bibinfo {author} {\bibfnamefont {Y.}~\bibnamefont
  {Wang}}\ and\ \bibinfo {author} {\bibfnamefont {S.}~\bibnamefont {Wang}},\
  }\href {\doibase 10.1103/PhysRevD.88.043522} {\bibfield  {journal} {\bibinfo
  {journal} {Phys.Rev.}\ }\textbf {\bibinfo {volume} {D88}},\ \bibinfo {pages}
  {043522} (\bibinfo {year} {2013})},\ \Eprint {http://arxiv.org/abs/1304.4514}
  {arXiv:1304.4514 [astro-ph.CO]} \BibitemShut {NoStop}%
\bibitem [{\citenamefont {Fixsen}(2009)}]{Fixsen:2009ug}%
  \BibitemOpen
  \bibfield  {author} {\bibinfo {author} {\bibfnamefont {D.}~\bibnamefont
  {Fixsen}},\ }\href {\doibase 10.1088/0004-637X/707/2/916} {\bibfield
  {journal} {\bibinfo  {journal} {Astrophys.J.}\ }\textbf {\bibinfo {volume}
  {707}},\ \bibinfo {pages} {916} (\bibinfo {year} {2009})},\ \Eprint
  {http://arxiv.org/abs/0911.1955} {arXiv:0911.1955 [astro-ph.CO]} \BibitemShut
  {NoStop}%
\bibitem [{\citenamefont {Hu}\ and\ \citenamefont
  {Sugiyama}(1996)}]{Hu:1995en}%
  \BibitemOpen
  \bibfield  {author} {\bibinfo {author} {\bibfnamefont {W.}~\bibnamefont
  {Hu}}\ and\ \bibinfo {author} {\bibfnamefont {N.}~\bibnamefont {Sugiyama}},\
  }\href {\doibase 10.1086/177989} {\bibfield  {journal} {\bibinfo  {journal}
  {Astrophys.J.}\ }\textbf {\bibinfo {volume} {471}},\ \bibinfo {pages} {542}
  (\bibinfo {year} {1996})},\ \Eprint {http://arxiv.org/abs/astro-ph/9510117}
  {arXiv:astro-ph/9510117 [astro-ph]} \BibitemShut {NoStop}%
\bibitem [{\citenamefont {Chuang}\ and\ \citenamefont
  {Wang}(2012)}]{ChuangWang2012}%
  \BibitemOpen
  \bibfield  {author} {\bibinfo {author} {\bibfnamefont {C.-H.}\ \bibnamefont
  {Chuang}}\ and\ \bibinfo {author} {\bibfnamefont {Y.}~\bibnamefont {Wang}},\
  }\href {\doibase 10.1111/j.1365-2966.2012.21565.x} {\bibfield  {journal}
  {\bibinfo  {journal} {Mon.Not.Roy.Astron.Soc.}\ }\textbf {\bibinfo {volume}
  {426}},\ \bibinfo {pages} {226} (\bibinfo {year} {2012})},\ \Eprint
  {http://arxiv.org/abs/1102.2251} {arXiv:1102.2251 [astro-ph.CO]} \BibitemShut
  {NoStop}%
\bibitem [{\citenamefont {Chuang}\ \emph {et~al.}(2013)\citenamefont {Chuang},
  \citenamefont {Prada}, \citenamefont {Cuesta}, \citenamefont {Eisenstein},
  \citenamefont {Kazin} \emph {et~al.}}]{Chuang:2013hya}%
  \BibitemOpen
  \bibfield  {author} {\bibinfo {author} {\bibfnamefont {C.-H.}\ \bibnamefont
  {Chuang}}, \bibinfo {author} {\bibfnamefont {F.}~\bibnamefont {Prada}},
  \bibinfo {author} {\bibfnamefont {A.~J.}\ \bibnamefont {Cuesta}}, \bibinfo
  {author} {\bibfnamefont {D.~J.}\ \bibnamefont {Eisenstein}}, \bibinfo
  {author} {\bibfnamefont {E.}~\bibnamefont {Kazin}},  \emph {et~al.},\
  }\href@noop {} {\  (\bibinfo {year} {2013})},\ \Eprint
  {http://arxiv.org/abs/1303.4486} {arXiv:1303.4486 [astro-ph.CO]} \BibitemShut
  {NoStop}%
\bibitem [{\citenamefont {Kazin}\ \emph {et~al.}(2010)\citenamefont {Kazin}
  \emph {et~al.}}]{Kazin:2009cj}%
  \BibitemOpen
  \bibfield  {author} {\bibinfo {author} {\bibfnamefont {E.~A.}\ \bibnamefont
  {Kazin}} \emph {et~al.} (\bibinfo {collaboration} {SDSS}),\ }\href {\doibase
  10.1088/0004-637X/710/2/1444} {\bibfield  {journal} {\bibinfo  {journal}
  {Astrophys. J.}\ }\textbf {\bibinfo {volume} {710}},\ \bibinfo {pages} {1444}
  (\bibinfo {year} {2010})},\ \Eprint {http://arxiv.org/abs/0908.2598}
  {arXiv:0908.2598 [astro-ph.CO]} \BibitemShut {NoStop}%
\bibitem [{\citenamefont {Eisenstein}\ \emph {et~al.}(2011)\citenamefont
  {Eisenstein} \emph {et~al.}}]{Eisenstein:2011sa}%
  \BibitemOpen
  \bibfield  {author} {\bibinfo {author} {\bibfnamefont {D.~J.}\ \bibnamefont
  {Eisenstein}} \emph {et~al.} (\bibinfo {collaboration} {SDSS}),\ }\href
  {\doibase 10.1088/0004-6256/142/3/72} {\bibfield  {journal} {\bibinfo
  {journal} {Astron. J.}\ }\textbf {\bibinfo {volume} {142}},\ \bibinfo {pages}
  {72} (\bibinfo {year} {2011})},\ \Eprint {http://arxiv.org/abs/1101.1529}
  {arXiv:1101.1529 [astro-ph.IM]} \BibitemShut {NoStop}%
\bibitem [{\citenamefont {Eisenstein}\ and\ \citenamefont
  {Hu}(1998)}]{Eisenstein:1997ik}%
  \BibitemOpen
  \bibfield  {author} {\bibinfo {author} {\bibfnamefont {D.~J.}\ \bibnamefont
  {Eisenstein}}\ and\ \bibinfo {author} {\bibfnamefont {W.}~\bibnamefont
  {Hu}},\ }\href {\doibase 10.1086/305424} {\bibfield  {journal} {\bibinfo
  {journal} {Astrophys.J.}\ }\textbf {\bibinfo {volume} {496}},\ \bibinfo
  {pages} {605} (\bibinfo {year} {1998})},\ \Eprint
  {http://arxiv.org/abs/astro-ph/9709112} {arXiv:astro-ph/9709112 [astro-ph]}
  \BibitemShut {NoStop}%
\bibitem [{\citenamefont {Suzuki}\ \emph {et~al.}(2012)\citenamefont {Suzuki},
  \citenamefont {Rubin}, \citenamefont {Lidman}, \citenamefont {Aldering},
  \citenamefont {Amanullah} \emph {et~al.}}]{Suzuki:2011hu}%
  \BibitemOpen
  \bibfield  {author} {\bibinfo {author} {\bibfnamefont {N.}~\bibnamefont
  {Suzuki}}, \bibinfo {author} {\bibfnamefont {D.}~\bibnamefont {Rubin}},
  \bibinfo {author} {\bibfnamefont {C.}~\bibnamefont {Lidman}}, \bibinfo
  {author} {\bibfnamefont {G.}~\bibnamefont {Aldering}}, \bibinfo {author}
  {\bibfnamefont {R.}~\bibnamefont {Amanullah}},  \emph {et~al.},\ }\href
  {\doibase 10.1088/0004-637X/746/1/85} {\bibfield  {journal} {\bibinfo
  {journal} {Astrophys.J.}\ }\textbf {\bibinfo {volume} {746}},\ \bibinfo
  {pages} {85} (\bibinfo {year} {2012})},\ \Eprint
  {http://arxiv.org/abs/1105.3470} {arXiv:1105.3470 [astro-ph.CO]} \BibitemShut
  {NoStop}%
\bibitem [{\citenamefont {Conley}\ \emph {et~al.}(2011)\citenamefont {Conley}
  \emph {et~al.}}]{Conley:2011ku}%
  \BibitemOpen
  \bibfield  {author} {\bibinfo {author} {\bibfnamefont {A.}~\bibnamefont
  {Conley}} \emph {et~al.} (\bibinfo {collaboration} {SNLS Collaboration}),\
  }\href {\doibase 10.1088/0067-0049/192/1/1} {\bibfield  {journal} {\bibinfo
  {journal} {Astrophys.J.Suppl.}\ }\textbf {\bibinfo {volume} {192}},\ \bibinfo
  {pages} {1} (\bibinfo {year} {2011})},\ \Eprint
  {http://arxiv.org/abs/1104.1443} {arXiv:1104.1443 [astro-ph.CO]} \BibitemShut
  {NoStop}%
\bibitem [{\citenamefont {Bennett}\ \emph {et~al.}(2014)\citenamefont
  {Bennett}, \citenamefont {Larson}, \citenamefont {Weiland},\ and\
  \citenamefont {Hinshaw}}]{Bennett:2014tka}%
  \BibitemOpen
  \bibfield  {author} {\bibinfo {author} {\bibfnamefont {C.}~\bibnamefont
  {Bennett}}, \bibinfo {author} {\bibfnamefont {D.}~\bibnamefont {Larson}},
  \bibinfo {author} {\bibfnamefont {J.}~\bibnamefont {Weiland}}, \ and\
  \bibinfo {author} {\bibfnamefont {G.}~\bibnamefont {Hinshaw}},\ }\href
  {\doibase 10.1088/0004-637X/794/2/135} {\bibfield  {journal} {\bibinfo
  {journal} {Astrophys.J.}\ }\textbf {\bibinfo {volume} {794}},\ \bibinfo
  {pages} {135} (\bibinfo {year} {2014})},\ \Eprint
  {http://arxiv.org/abs/1406.1718} {arXiv:1406.1718 [astro-ph.CO]} \BibitemShut
  {NoStop}%
\bibitem [{\citenamefont {Christensen}\ \emph {et~al.}(2001)\citenamefont
  {Christensen}, \citenamefont {Meyer}, \citenamefont {Knox},\ and\
  \citenamefont {Luey}}]{Christensen:2001gj}%
  \BibitemOpen
  \bibfield  {author} {\bibinfo {author} {\bibfnamefont {N.}~\bibnamefont
  {Christensen}}, \bibinfo {author} {\bibfnamefont {R.}~\bibnamefont {Meyer}},
  \bibinfo {author} {\bibfnamefont {L.}~\bibnamefont {Knox}}, \ and\ \bibinfo
  {author} {\bibfnamefont {B.}~\bibnamefont {Luey}},\ }\href {\doibase
  10.1088/0264-9381/18/14/306} {\bibfield  {journal} {\bibinfo  {journal}
  {Class.Quant.Grav.}\ }\textbf {\bibinfo {volume} {18}},\ \bibinfo {pages}
  {2677} (\bibinfo {year} {2001})},\ \Eprint
  {http://arxiv.org/abs/astro-ph/0103134} {arXiv:astro-ph/0103134 [astro-ph]}
  \BibitemShut {NoStop}%
\bibitem [{\citenamefont {Lewis}\ and\ \citenamefont
  {Bridle}(2002)}]{Lewis:2002ah}%
  \BibitemOpen
  \bibfield  {author} {\bibinfo {author} {\bibfnamefont {A.}~\bibnamefont
  {Lewis}}\ and\ \bibinfo {author} {\bibfnamefont {S.}~\bibnamefont {Bridle}},\
  }\href {\doibase 10.1103/PhysRevD.66.103511} {\bibfield  {journal} {\bibinfo
  {journal} {Phys.Rev.}\ }\textbf {\bibinfo {volume} {D66}},\ \bibinfo {pages}
  {103511} (\bibinfo {year} {2002})},\ \Eprint
  {http://arxiv.org/abs/astro-ph/0205436} {arXiv:astro-ph/0205436 [astro-ph]}
  \BibitemShut {NoStop}%
\bibitem [{\citenamefont {Trotta}\ and\ \citenamefont
  {Durrer}(2004)}]{Trotta:2004qj}%
  \BibitemOpen
  \bibfield  {author} {\bibinfo {author} {\bibfnamefont {R.}~\bibnamefont
  {Trotta}}\ and\ \bibinfo {author} {\bibfnamefont {R.}~\bibnamefont
  {Durrer}},\ }\href@noop {} {\  (\bibinfo {year} {2004})},\ \Eprint
  {http://arxiv.org/abs/astro-ph/0410115} {arXiv:astro-ph/0410115 [astro-ph]}
  \BibitemShut {NoStop}%
\bibitem [{\citenamefont {Dunkley}\ \emph {et~al.}(2005)\citenamefont
  {Dunkley}, \citenamefont {Bucher}, \citenamefont {Ferreira}, \citenamefont
  {Moodley},\ and\ \citenamefont {Skordis}}]{Dunkley2005}%
  \BibitemOpen
  \bibfield  {author} {\bibinfo {author} {\bibfnamefont {J.}~\bibnamefont
  {Dunkley}}, \bibinfo {author} {\bibfnamefont {M.}~\bibnamefont {Bucher}},
  \bibinfo {author} {\bibfnamefont {P.~G.}\ \bibnamefont {Ferreira}}, \bibinfo
  {author} {\bibfnamefont {K.}~\bibnamefont {Moodley}}, \ and\ \bibinfo
  {author} {\bibfnamefont {C.}~\bibnamefont {Skordis}},\ }\href {\doibase
  10.1111/j.1365-2966.2004.08464.x/abs/} {\bibfield  {journal} {\bibinfo
  {journal} {Mon.Not.Roy.Astron.Soc.}\ }\textbf {\bibinfo {volume} {356}},\
  \bibinfo {pages} {925} (\bibinfo {year} {2005})},\ \Eprint
  {http://arxiv.org/abs/astro-ph/0405462} {arXiv:astro-ph/0405462 [astro-ph]}
  \BibitemShut {NoStop}%
\bibitem [{\citenamefont {Carroll}\ \emph {et~al.}(1992)\citenamefont
  {Carroll}, \citenamefont {Press},\ and\ \citenamefont
  {Turner}}]{Carroll:1991mt}%
  \BibitemOpen
  \bibfield  {author} {\bibinfo {author} {\bibfnamefont {S.~M.}\ \bibnamefont
  {Carroll}}, \bibinfo {author} {\bibfnamefont {W.~H.}\ \bibnamefont {Press}},
  \ and\ \bibinfo {author} {\bibfnamefont {E.~L.}\ \bibnamefont {Turner}},\
  }\href {\doibase 10.1146/annurev.aa.30.090192.002435} {\bibfield  {journal}
  {\bibinfo  {journal} {Ann.Rev.Astron.Astrophys.}\ }\textbf {\bibinfo {volume}
  {30}},\ \bibinfo {pages} {499} (\bibinfo {year} {1992})}\BibitemShut
  {NoStop}%
\bibitem [{\citenamefont {Sahni}\ and\ \citenamefont
  {Starobinsky}(2000)}]{Sahni:1999gb}%
  \BibitemOpen
  \bibfield  {author} {\bibinfo {author} {\bibfnamefont {V.}~\bibnamefont
  {Sahni}}\ and\ \bibinfo {author} {\bibfnamefont {A.~A.}\ \bibnamefont
  {Starobinsky}},\ }\href@noop {} {\bibfield  {journal} {\bibinfo  {journal}
  {Int.J.Mod.Phys.}\ }\textbf {\bibinfo {volume} {D9}},\ \bibinfo {pages} {373}
  (\bibinfo {year} {2000})},\ \Eprint {http://arxiv.org/abs/astro-ph/9904398}
  {arXiv:astro-ph/9904398 [astro-ph]} \BibitemShut {NoStop}%
\bibitem [{\citenamefont {Knop}\ \emph {et~al.}(2003)\citenamefont {Knop} \emph
  {et~al.}}]{Knop:2003iy}%
  \BibitemOpen
  \bibfield  {author} {\bibinfo {author} {\bibfnamefont {R.~A.}\ \bibnamefont
  {Knop}} \emph {et~al.} (\bibinfo {collaboration} {Supernova Cosmology
  Project}),\ }\href {\doibase 10.1086/378560} {\bibfield  {journal} {\bibinfo
  {journal} {Astrophys.J.}\ }\textbf {\bibinfo {volume} {598}},\ \bibinfo
  {pages} {102} (\bibinfo {year} {2003})},\ \Eprint
  {http://arxiv.org/abs/astro-ph/0309368} {arXiv:astro-ph/0309368 [astro-ph]}
  \BibitemShut {NoStop}%
\bibitem [{\citenamefont {Riess}\ \emph {et~al.}(2004)\citenamefont {Riess}
  \emph {et~al.}}]{Riess:2004nr}%
  \BibitemOpen
  \bibfield  {author} {\bibinfo {author} {\bibfnamefont {A.~G.}\ \bibnamefont
  {Riess}} \emph {et~al.} (\bibinfo {collaboration} {Supernova Search Team}),\
  }\href {\doibase 10.1086/383612} {\bibfield  {journal} {\bibinfo  {journal}
  {Astrophys.J.}\ }\textbf {\bibinfo {volume} {607}},\ \bibinfo {pages} {665}
  (\bibinfo {year} {2004})},\ \Eprint {http://arxiv.org/abs/astro-ph/0402512}
  {arXiv:astro-ph/0402512 [astro-ph]} \BibitemShut {NoStop}%
\bibitem [{\citenamefont {Mukherjee}\ \emph {et~al.}(2006)\citenamefont
  {Mukherjee}, \citenamefont {Parkinson},\ and\ \citenamefont
  {Liddle}}]{Mukherjee:2005wg}%
  \BibitemOpen
  \bibfield  {author} {\bibinfo {author} {\bibfnamefont {P.}~\bibnamefont
  {Mukherjee}}, \bibinfo {author} {\bibfnamefont {D.}~\bibnamefont
  {Parkinson}}, \ and\ \bibinfo {author} {\bibfnamefont {A.~R.}\ \bibnamefont
  {Liddle}},\ }\href {\doibase 10.1086/501068} {\bibfield  {journal} {\bibinfo
  {journal} {Astrophys.J.}\ }\textbf {\bibinfo {volume} {638}},\ \bibinfo
  {pages} {L51} (\bibinfo {year} {2006})},\ \Eprint
  {http://arxiv.org/abs/astro-ph/0508461} {arXiv:astro-ph/0508461 [astro-ph]}
  \BibitemShut {NoStop}%
\bibitem [{\citenamefont {Trotta}(2007)}]{Trotta:2005ar}%
  \BibitemOpen
  \bibfield  {author} {\bibinfo {author} {\bibfnamefont {R.}~\bibnamefont
  {Trotta}},\ }\href {\doibase 10.1111/j.1365-2966.2007.11738.x} {\bibfield
  {journal} {\bibinfo  {journal} {Mon.Not.Roy.Astron.Soc.}\ }\textbf {\bibinfo
  {volume} {378}},\ \bibinfo {pages} {72} (\bibinfo {year} {2007})},\ \Eprint
  {http://arxiv.org/abs/astro-ph/0504022} {arXiv:astro-ph/0504022 [astro-ph]}
  \BibitemShut {NoStop}%
\bibitem [{\citenamefont {Gordon}\ and\ \citenamefont
  {Trotta}(2007)}]{Gordon:2007xm}%
  \BibitemOpen
  \bibfield  {author} {\bibinfo {author} {\bibfnamefont {C.}~\bibnamefont
  {Gordon}}\ and\ \bibinfo {author} {\bibfnamefont {R.}~\bibnamefont
  {Trotta}},\ }\href {\doibase 10.1111/j.1365-2966.2007.12707.x} {\bibfield
  {journal} {\bibinfo  {journal} {Mon.Not.Roy.Astron.Soc.}\ }\textbf {\bibinfo
  {volume} {382}},\ \bibinfo {pages} {1859} (\bibinfo {year} {2007})},\ \Eprint
  {http://arxiv.org/abs/0706.3014} {arXiv:0706.3014 [astro-ph]} \BibitemShut
  {NoStop}%
\end{thebibliography}%

\end{document}